# Assessing the global and local uncertainty in scientific evidence in the presence of model misspecification

**Mark L. Taper[1,3]\***, **Subhash R. Lele[2]**, **José M. Ponciano[3]**, **Brian Dennis[4]**, and **Christopher L. Jerde[5]**

[1]Montana State University, Department of Ecology, Bozeman MT, 59717, USA.
[2]University of Alberta, Department of Mathematical and Statistical Sciences, Edmonton, AB, T6G2G1, Canada
[3]University of Florida, Department of Biology, Gainesville FL, 32611-8525, USA.
[4]University of Idaho, Department of Fish and Wildlife Sciences and Department of Mathematics and Statistical Science, Moscow ID 83844-1136, USA
[5] Marine Science Institute, University of California, Santa Barbara, CA 93106-6150, USA

**\* Correspondence:**
Mark L. Taper
MarkLTaper@gmail.com

### Abstract

Scientists need to compare the support for models based on observed phenomena. The main goal of the evidential paradigm is to quantify the strength of evidence in the data for a reference model relative to an alternative model. This is done via an evidence function, such as $\Delta SIC$, an estimator of the sample size scaled difference of divergences between the generating mechanism and the competing models. To use evidence, either for decision making or as a guide to the accumulation of knowledge, an understanding of the uncertainty in the evidence is needed. This uncertainty is well characterized by the standard statistical theory of estimation. Unfortunately, the standard theory breaks down if the models are misspecified, as it is normally the case in scientific studies. We develop non-parametric bootstrap methodologies for estimating the sampling distribution of the evidence estimator under model misspecification. This sampling distribution allows us to determine how secure we are in our evidential statement. We characterize this uncertainty in the strength of evidence with two different types of confidence intervals, which we term "global" and "local". We discuss how evidence uncertainty can be used to improve scientific inference and illustrate this with a reanalysis of the model identification problem in a prominent landscape ecology study (Grace and Keeley, 2006) using structural equations.

# 1    Introduction

> *When a person supposes that he knows, and does not know; this appears to be the great source of all the errors of the intellect.*
>
> Plato. The Sophist. 360 B.C.E Translated by Benjamin Jowett

One of the main goals of scientific inference is to delineate and understand the underlying mechanism of a phenomenon of interest. In practice, we have several different hypotheses or proposed mechanisms, and we want to use the observed data to quantify the strength of evidence for one mechanism over thes alternative. The evidential approach to statistical and scientific inference uses estimates of the difference of the divergences from the true mechanism to the competing mechanisms to quantify the strength of evidence in the observed data for one mechanism over the other. The evidence function is an estimator of the sample size scaled divergence difference between two candidate statistical mechanisms (Lele, 2004). Importantly, evidence functions can be applied pairwise to multiple models to determine the support for multiple alternative mechanisms.

Various papers (Lele 2004; Taper and Lele 2011; Taper and Ponciano 2016; Jerde et al., 2019) discuss the desiderata that an evidence function should satisfy. In comparing a reference model to an alternative, the log-likelihood ratio (LLR) is the most used evidence function that is based on the Kullback-Leibler (KL) divergence. An evidence function is usually constructed so that if the realized value of the evidence function, the observed evidence, is larger than a pre-specified positive threshold value $(k_R)$, we say that data strongly support the reference model. If it is below a negative threshold value $(k_A)$ (i.e. closer to negative infinity), data strongly support the alternative model. If the evidence function is in between these two thresholds, data are said to be unable to distinguish between the two models.

A commonly used alternative to the evidential framework, Neyman-Pearson tests accords a special statistical status to the null model in that the type I error probability is fixed (does not depend on sample size) and the P-value is calculated with only the null model. Consequently, a variety of inferential distortions can famously occur when Neyman-Pearson testing is used for purposes beyond its working specifications (Dennis et al., 2019). By contrast, in the evidential framework no special status is accorded either the reference or alternative model. The designations of reference or alternative serve only to help an analyst understand which model is supported (relative to the other) by positive or negative evidence and do not confer any differences in statistical properties. Royall (1997, 2000) considers the situation where the reference and alternative models are fully specified, that is, there are no unknown parameters that need to be estimated from the data. Under the assumption the reference model is the true generating mechanism, he uses the asymptotic distribution of the LLR to compute the probability of misleading evidence, namely the probability that observed evidence would strongly support the alternative mechanism. He also considers the probability of weak evidence, namely the probability of being unable to distinguish between the two mechanisms. Following the results of Godambe (1960), Lele (2004) shows that, under regularity conditions, among all evidence functions the LLR is optimal in the sense that the rate at which the probability of strong evidence converges to 1 is the fastest. These error probabilities, especially the probability of weak evidence, are useful for pre-experiment decisions on sample size (Strug et al. 2007) or optimal designing of experiments.

Dennis et al. (2019) recognize the reality that most models are only approximations and hence the true generating mechanism is likely to be neither the reference nor the alternative modelFollowing Dennis et al. (2019), we consider a model misspecified if the data distribution it predicts cannot be made to match the distribution of the true generating process by appropriate parameterization. A model set is misspecified if all of its members are misspecified. In practice, the model sets used in science are almost always misspecified to some degree and may be badly misspecified particularly during early exploration of scientific phenomena.

The asymptotic distribution of the LLR under model misspecification (Dennis et al., 2019, Vuong, 1989) depends on the geometry of the misspecification, that is, how the true generating mechanism and the two competing model spaces relate to each other. . In scientific studies, instead of fully specified reference and alternative models, one generally has reference and alternative model spaces, a set of parametric models whose parameters need to be estimated using the observed data. Such a set forms a space because its elements have geometrical relationships such as divergences between them. Dennis et al. (2019) use the asymptotic distribution of the LLR to compute the error probabilities in comparing model spaces when the true generating model might be outside the specified model spaces. In Table 3, we list all possible topologies, i.e. configurations, for the generating mechanism and competing model spaces and corresponding asymptotic distributions of the LLR. One important feature of these asymptotic distributions is the means of these distributions increase toward infinity at rates proportional to sample size, $n$, whereas the standard deviations increase toward infinity at rates proportional to $n^{1/2}$, producing tail probabilities (probabilities of misleading evidence) that converge to zero (because the coefficient of variation goes to zero). Thus, in all evidential comparisons using the LLR, as the sample size increases, probability of strong evidence for the best approximating mechanism converges to 1 and all other error probabilities converge to 0 (Dennis et al., 2019).

As discussed by Royall (1997, 2000, 2004), this behavior of the error probabilities is in stark contrast to the classical Neyman-Pearson approach where the probability of type I error remains constant for all sample sizes. The consequence to the applied scientist is that the true generating mechanism is rejected in favor of a misspecified null some fraction of the time regardless of the amount of data collected. Of course, classical statistical inference does not stop at hypothesis testing. It also computes the sampling distribution of the estimator of the effect size. Unlike the probability of type I error in hypothesis testing, as the sample size increases, the sampling distribution does concentrate around the true effect size, thus leading to the correct inference. Royall (2000) and Dennis et al. (2019) obtain this sampling distribution asymptotically.

The goal of this paper is to obtain a fuller understanding of uncertainty in observed evidence under realistic sample sizes by estimating the finite sample sampling distribution of the strength of evidence under model misspecification via non-parametric bootstrap. In an earlier paper, Taper and Lele (2011) had suggested the use of non-parametric bootstrap to understand finite sample uncertainty in observed evidence when the true generating mechanism may be different than the reference and alternative models. This current paper is a detailed exploration of this suggestion.

The non-parametric bootstrap is a computational approach (Efron and Tibshirani 1993) used to get a finite sample approximation to the sampling distribution of a statistic that is valid under model misspecification. Generally, the sampling distribution of the estimator is far more useful for supporting scientific arguments than is a hypothesis test by itself (Xie and Singh, 2013; Schweder, 2018). An inferential statement is any statement about the parameters, form of the underlying mechanism, or a future outcome. An inferential statement becomes a statistical inferential statement only when a measure of uncertainty is attached to it (Cox 1958). An accessible review of various approaches to quantifying uncertainty in an inferential statement is available in Lele (2020). The classical frequentist inference uses aleatory probability (frequency of an event under hypothetical infinite replication of experiment) to quantify uncertainty of an inferential statement. To obtain the aleatory uncertainty of an inferential statement, a critical question that needs to be answered is: which experiment/sampling design do we (hypothetically) repeat? Lele (2020) uses the simple linear regression model to illustrate the distinction between the global (also known as unconditional, pre-data or, pre-experiment) and local (also known as conditional, post-data or, post-experiment) uncertainty. In this paper, we augment that illustration by comparing the differences between global and local uncertainty in mark-recapture analysis (see box 2.)

We consider the problem of model selection. As was described in Dennis et al. (2019), the reference and the alternative models are not fully specified. There are unknown parameters that need to be estimated and hence the set-up discussed in Royall (1997) does not apply. Because these two competing models may involve different number of parameters, an unmodified LLR is not an appropriate evidence function, and the LLR needs to be penalized for the number of parameters to be estimated (Akaike, 1973). Furthermore, to make the error probabilities of misleading and weak evidence to converge to 0 as sample size increases, we also need to moderate the penalty by a function of the sample size that grows to infinity at a rate between $\log(\log(n))$ and $n$ (Nishii, 1988). The appropriate evidence functions for the model selection problem are based on the consistent information criteria such as the Schwarz's Information Criterion (SIC) (Schwarz, 1978) that incorporates both the sample size and the number of parameters in its penalty term. Inconsistent criteria, such as the Akaike Information Criterion (AIC), tend to overfit at all sample sizes and do not lead to valid evidence functions due to the absence of an augmentation of the penalty by the sample size. Note that despite having a sample size correction, the AICc (Hurvich and Tsai, 1989) is not consistent. Its sample size correction is aimed at correcting small sample bias, not large sample inconsistency. We will return to this point in the discussion. One can potentially use any divergence measure other than the Kullback-Leibler divergence and with appropriate (i.e. consistent) sample size and parameter size penalty function, one can create a valid evidence function. The evidence function is, in fact, a *scaled and penalized difference between the estimates of divergences of two models each to the generating process*.

In this paper, we show that model selection based on a bootstrap bias corrected information criterion known as the extended information criterion (EIC) (e.g. Kitagawa and Konishi, 2010) is strongly connected to various bias corrections of the profile likelihood (e.g. Pace and Salvan, 2006). We combine these two ideas with the use of a consistent penalty and show that a non-parametric bootstrap approach can be used to obtain finite sample and consistent, global and local uncertainty in the observed strength of evidence for the reference vis a vis the alternative model. The mathematical details are given in section 4. As a consequence of this development, we will use as our evidence function the mean of a bootstrapped distribution of $\Delta \text{SIC}$ s. However, while Pace and Salvan (2006) and Kitagawa and Konishi (2010) use the bootstrap only for computing the bias correction factor, we also use the entire sampling distribution to obtain valid, finite sample, global and local confidence intervals for the strength of evidence. That is, our confidence intervals will also be based on the quantiles of a bootstrapped distribution of $\Delta \text{SIC}$ s.

These confidence intervals are useful to make scientific conclusions. For example, if most of the sampling distribution is above the threshold, we have not only strong evidence, but it is also very unlikely to be strong by chance. We define such evidence as secure. If the sampling distribution is such that substantial portion is below the threshold, the observed evidence may be strong, but it cannot be considered secure and more data may be needed to clarify the situation.

Hoping to stimulate practicing scientists with the utility of our approach before they encounter the mathematics of our methods, this paper proceeds as follows: In section 2, we discuss the implications of uncertainty in evidence and the use of sampling distribution of the strength of evidence in drawing scientific conclusions in detail. In section 3 we apply these ideas in a reanalysis of a prominent ecological experiment analyzed using structural equations models (SEM) and discuss the scientific implications of the uncertainty in the strength of evidence. Section 4 describes the underlying mathematical concepts and the methodology for computing finite sample, global and local sampling distributions of the strength of evidence for model selection. In section 5, we validate the methodology using simulations for model selection in linear regression. In the last section, section 6, we discuss implications of the uncertainty quantification of the strength of evidence for the pursuance of science and suggest avenues for further research.

## 2    Scientific inference under evidential uncertainty

First, we note that simulations as well as the analytical results in Dennis et al. (2019) show that the sampling variability in evidence can be substantial. Hence using empirical evidence without a measure of uncertainty can be dangerous in practice leading to overconfidence, wrong decisions, misleading inferences, and misguided scientific enquiry. Furthermore, under model misspecification, evidence functions, such as the LLR and others become detached from model-based estimates of error probabilities and are just measures of relative plausibility (Barnard, 1949; Fisher, 1960; Sprott, 2000). Non-parametric confidence intervals on the strength of inference allow us to reattach our inferences to probability measures, although there is a considerable difference in what those probabilities mean between global and local inference. Before discussing the methodology to quantify global and local uncertainties in evidence and their real-world applications, let us first discuss how the sampling distribution of the strength of evidence could be used to draw scientific conclusions.

Royall (1997) considers three categories of strength of evidence: Strong evidence for a reference model, Strong evidence for the alternative model, and Weak evidence when the strength of evidence cannot distinguish between the two models. Often in ecological analysis, one finds the strength of evidence that is neither so weak that one feels comfortable saying one cannot distinguish between the models nor so strong that one is willing to stake a reputation on it. Hence, we suggest using five categories for strength of evidence, inserting categories of Prognostic evidence for the reference model and Prognostic evidence for the alternative. See Box 1 for a more complete discussion.

---

Box 1 Categories of strength of evidence

Often in ecological analysis, one finds evidence that is neither so weak that one feels comfortable saying one cannot distinguish between the models at all nor so strong that one is willing to stake a reputation on it. Thus, to the thresholds $k_A$ and $k_R$ we add the thresholds $k_a$ and $k_r$. Evidence between the thresholds $k_A$ and $k_a$ and between $k_r$ and $k_R$ could reasonably be called moderate, but to avoid a clash in abbreviations with the error category of misleading evidence, we will call such evidence prognostic. Now evidence is divided into five categories: strong evidence for the alternative model, prognostic evidence for the alternative model, evidence so weak that it is best to say that neither model is favored, prognostic evidence for the reference model, and strong evidence for the reference model.

1.  Strong evidence for the reference model if the strength of evidence is larger than $k_R$.

2.  Prognostic evidence for the reference model if the strength of evidence is between $k_r$ and $k_R$

3.  Weak evidence favoring neither model if the strength of evidence is between $k_a$ and $k_r$.

4.  Prognostic evidence for the alternative model if the strength of evidence is between $k_A$ and $k_a$

5.  Strong evidence for the alternative model if the strength of evidence is less than $k_A$

Royall (1997) pointed out that on occasion, one can have strong evidence that one model, say the reference, in your comparison is closer to the generating process than the other, say the alternative, when in fact it is the alternative that is truly closer to the generating process. Royall called such counterfactual evidence 'misleading'. With the weaker category of prognostic evidence, it is even more likely that evidence that is counterfactual will be estimated. We designate counterfactual prognostic evidence as "confusing evidence". With real data, one does not know if strong evidence is in fact misleading, or if prognostic evidence is confusing. However, in design and validation studies, whether analytic or computational, the researcher does know when evidence is misleading or confusing, and these categories are very helpful (see section 5).

It is important to realize that the sign of evidence only indicates which model is estimated to be closer to the generating process, positive for the reference model and negative for the alternative. Previously in the literature, $k_A$ has been set symmetrically to $-k_R$. In specific cases, there could be reason for asymmetry in thresholds, either because of asymmetry in probability models or because of decision cost. For simplicity, we adopt symmetric thresholds with $-k_p$ and $k_p$ indicating the thresholds between weak evidence and prognostic evidence for the alternative and reference models respectively. Similarly, $-k_S$ and $k_S$ are the thresholds between prognostic evidence and strong evidence for the alternative and reference models. Jerde et al. (2019) discuss interpretations for levels of evidence. Following their recommendations, we define $k_p \equiv 4$ and $k_S \equiv 7$.

____________________End of Box 1____________________________________________

One final difference between Royall's characterization of the strength of evidence and our characterization is that Royall considered the strength of evidence a ratio of likelihoods. We, on the other hand always consider strength of evidence as differences on a logarithmic scale (see discussion in Barnard, 1949). This ties our conceptualization more closely with information theory and the comparison of divergences.

## 2.1   Understanding global and local uncertainty in evidence

Confidence intervals are a mainstay in ecological inference, increasingly and justifiably so (Johnson, 1999; Ponciano et al., 2009; Halsey, 2019; Holland, 2019; Fieberg et al., 2020). They transmit a more complete and interpretable representation of the information in data than do hypothesis tests. A confidence interval is a range of values for a statistic, a function of the data, that is expected to cover (capture, include) an estimation target a given per cent of the time (e.g. 95%) under repetition of a specified hypothetical experiment (Neyman, 1937). The target of an interval is something in nature about which we would like to make an inference such as a population parameter or a function of a parameter.

For evidence, there are both local and global intervals that can be calculated (see section 4 for details). In order to understand confidence intervals for evidence, it is important to realize that not only are the interval widths different, but that the targets are also different.

The global target is the difference between the divergences of the best possible representations of the two models to the natural generating process. The uncertainty in the global interval includes the sampling uncertainty for the data, model estimation uncertainty given the data, and uncertainty due to model set misspecification.

The local target is the evidence *in the observed data* for the best possible representation of one model over the best possible representation of the other model. The uncertainty in the global interval represents just the model estimation uncertainties given the observed data, and uncertainty due to model set misspecification.

Global intervals reflect the variation in the estimates if independent experiments are conducted in a manner like the original experiment. The local intervals reflect the informativeness of the specific experimental outcome in hand.

The local interval can capitalize on lucky samples to make precise inferences about the strength of evidence for the reference model relative to the alternative model. On the other hand, with unlucky samples where the parameter estimate may be far from the truth, the local intervals also end up making precise but misleading inferential statements. Global intervals, because they average over all possible datasets, tend to be wider than the local intervals. They are conservative in their uncertainty quantification, making strong inferential statements only cautiously. That does not mean that the global intervals are without use. Scientific results need to be validated by independent replication. A global interval indicates how discrepant the results of a repetition of the experiments could be from the original before contradicting your results and hence protects against the possibility of being contradicted.

---

Box 2 Global and local intervals in Mark/Recapture analysis

In ecology, where uncertainty in the study systems is ubiquitous, it is common practice to formulate a scientific hypothesis in the form of a simplified probabilistic model of how the data arose. This simplification allows the analysts to focus the inferential process on a typically small set of quantities bearing strong ecological or management importance. Such simplifications are in fact conceptual restrictions on how the data arose and are used to formulate the likelihood function. Multiple uncertainty simplifications/restrictions are incorporated in the form of multiple conditioning layers. Take for instance a simple closed population mark-recapture experiment where in a first visit to a study area, a number of animals of the species of interest are marked and released. In a second visit, a sample of animals from the same population are captured and the number of previously marked animals in that sample recorded. Under that setting, different levels of conditioning restrict more and more the sampling uncertainty while keeping the focus on the same inferential quantity of interest – the total population size. We prefer the terms 'global' and 'local' because they evoke the scope of inference that can be addressed by each type of uncertainty. The sampling distribution for global uncertainty is computed using the entire sample space whereas 'local' uncertainty is computed using a relevant subset of the sample space (Buehler, 1959).

The key question in global and local inference is what components of your data do you want to be considered fixed (or given) and what components do you want to be considered random (or representative). A completely unconstrained interval is considered global. Intervals with constraints are considered local. An alternative way of approaching this question, which may be clearer for some, is to recognize that a confidence interval represents the variability in hypothetically repeated experiments. When you treat a component as fixed or random, you are specifying different hypothetical experiments. One of the goals of confidence intervals is to define what estimates a skeptic who tries to replicate the experiment might obtain. Different types of experimental conditions that the skeptic might use dictate the choice of the interval.

We illustrate the concepts of global and local inference using the familiar problem of population size estimation using the Lincoln-Peterson estimator. We use the data from a published experiment on iguana population density to create a realistic framework along with some R commands to demark the global and local differences clearly in the calculations. The data and a more complete treatment can be found in Powell and Gale (2015).

Below is a mark-recapture data set, describing one re-sampling occasion. On day 3 of their experiment 131 individuals, n, are captured and 116, x, of these have previously been marked. Initially (days 0,1 and 2) m=221 individuals have been captured, marked and released:

```
m <- 221
n <- 131
x <- 116
```

From these data we estimate a total population size using the Lincoln-Petersen estimator. Thus the target for point and interval estimation is the true population size. As it happens, the same estimator is obtained whether you assume that:  1) Bo'h''m' a'd''n' are fixed. ')''m' is considered fixed, b't''n' is not.  And 3) Bot','m' a'd''n' are considered random.

While the estimate of the total population for these three cases is identical, the uncertainty around it is not. Each set of assumptions fully determines the confidence intervals.  We demonstrate this via Parametric Bootstrap (PB) because of how the levels of randomness enter at each stage is much more perspicuous in the PB code than in the corresponding analytic formulae.

Parametric Bootstrap

Compute the Lincoln-Petersen estimator for the sample at hand as well as the nuisance parameter phi.hat (the capture probability)

```
t.hat <- floor((n*m)/x)
print(t.hat)

## [1] 249

phi.hat <- n/t.hat # estimated capture probability
print(phi.hat)

## [1] 0.5261044
```

Now let's set our PB simulation parameters to these two estimates:

```
t.true <- t.hat
phi.true <- phi.hat
```

Next, set the total number of simulations

```
B <- 10000
```

and then create empty arrays to store the three types of estimates

```
# Lincoln Petersen constrained on m and n (ultimate local: fix'd''m' a'd''n')
LP.mn.bt <- rep(NA,B)

#Lincoln Petersen constrained on m (local-fixed- m, but glob'l''n': rand'm''N')
LP.m.bt <- rep(NA,B)

#Lincoln Petersen unconstrained (Glob'l''m' and glob'l''n'. Both are random: M &
N)
LP.bt <- rep(NA,B)
```

Finally, just turn the crank on the PB iterations and store them:

```
fIr(® in 1:B){

  #### Simulating data and computi'g 't.'at' under the first assumption:
  X.mn <- rhyper(nn=1, m=m,n=(t.true-m),k=n) #constrained on m and n
  LP.mn.bt[i] <- m*n/X.mn

  #### Simulating data and computi'g 't.'at' under the second assumption
  N <- rbinom(n=1,size=t.true,prob=phi.true) # unconstrained
  X.m <- rbinom(n=1, size=min(m,N), prob=m/t.true)  #constrained on m but not n
  LP.m.bt[i] <-  m*N/X.m

  #### Simulating data and computi'g 't.'at' under the third assumption
  M <- rbinom(n=1,size=t.true,prob=phi.true) # unconstrained
  X <- rbinom(n=1, size=min(M,N), prob=M/t.true)     # not constrained on either
m or n
  LP.bt[i]   <-  M*N/X
}

# Throw out the outcomes for which x=0. A result of x=0 is possible, but gives
# an infinite estimate of population size.
LP.mn.bt <- LP.mn.bt[is.finite(LP.mn.bt)]
LP.m.bt <- LP.m.bt[is.finite(LP.m.bt)]
LP.bt <- LP.bt[is.finite(LP.bt)]
```

It is instructive to look at the sample spaces for these three estimators:

Sample Spaces:

LP.bt: $\Omega_G = \left\{ M \in \{0, \cdots, T\}, N \in \{0, \cdots, T\}, X \left\{ \max\left(0, N - (T-M)\right), \cdots, \min(M, N)\right\} \right\}$

LP.m.bt: $\Omega_{L_1} = \left\{ m, N \in \{0, \cdots, T\}, X \left\{ \max\left(0, N - (T-m)\right), \cdots, \min(m, N)\right\} \right\}$

LP.mn.bt: $\Omega_{L_2} = \left\{ m, n, X \left\{ \max\left(0, n - (T-m)\right), \cdots, \min(m, n)\right\} \right\}$,

where $T$ is the true population size.

The sample spaces are all possible data sets that the simulations could generate under each of the model assumptions. The sample space for LP.m.bt is nested within that of LP.m.bt, which is itself nested within the sample space of LP.bt. Clearly, global and local are relative terms. LP.m.bt is local with respect to LP.bt, but global with respect to LP.mn.bt.

The sampling distributions for the three estimators are plotted in the figure below. We now have three different confidence intervals. Which is right? Statistics by itself cannot answer that question. These three intervals represent the uncertainty in the hypothetical repetition of three different experiments. In the type 1 experiment, with $m$ and $n$ constrained, the only thing that can vary experiment to experiment is the number of marked animals in the final day sample.

In type 2, the number of previously marked individuals is constrained but not the final day Fsample size. The hypothetical experiment is repeated only for the final day; varying numbers of individuals as well as varying numbers of marked animals may be captured on the final day. In type 3, the entire hypothetical experiment is repeated. The number of marked individuals, the number of captured individuals, and the number of marked individuals in the second sample may all vary.

The appropriate interval depends on the kind of uncertainty you are trying to represent. The first interval answers the question: How different the estimators of the total population could be if someone else replicated the experiment such that the total number of marked individuals and total number of captures are identical to your experiment? This can happen in a field survey where the total number of marked animals and total number of captures is fixed by design, a priori. These numbers may depend on the budget the researcher might have for capturing animals for marking and for recapturing.

In the situations, such as camera trap surveys, the total number of marked animals may be fixed by design but the total number of captures, by the nature of the survey technique, is random. The second interval considers this possibility and allows for the randomness in the number of captures to compute the uncertainty in the total population size estimator.

In the case of fish surveys, the number of fish caught in the traps or by electrofishing for marking is necessarily random and so is the number of fish in the sample afterwards. In this case, the third interval will be appropriate.

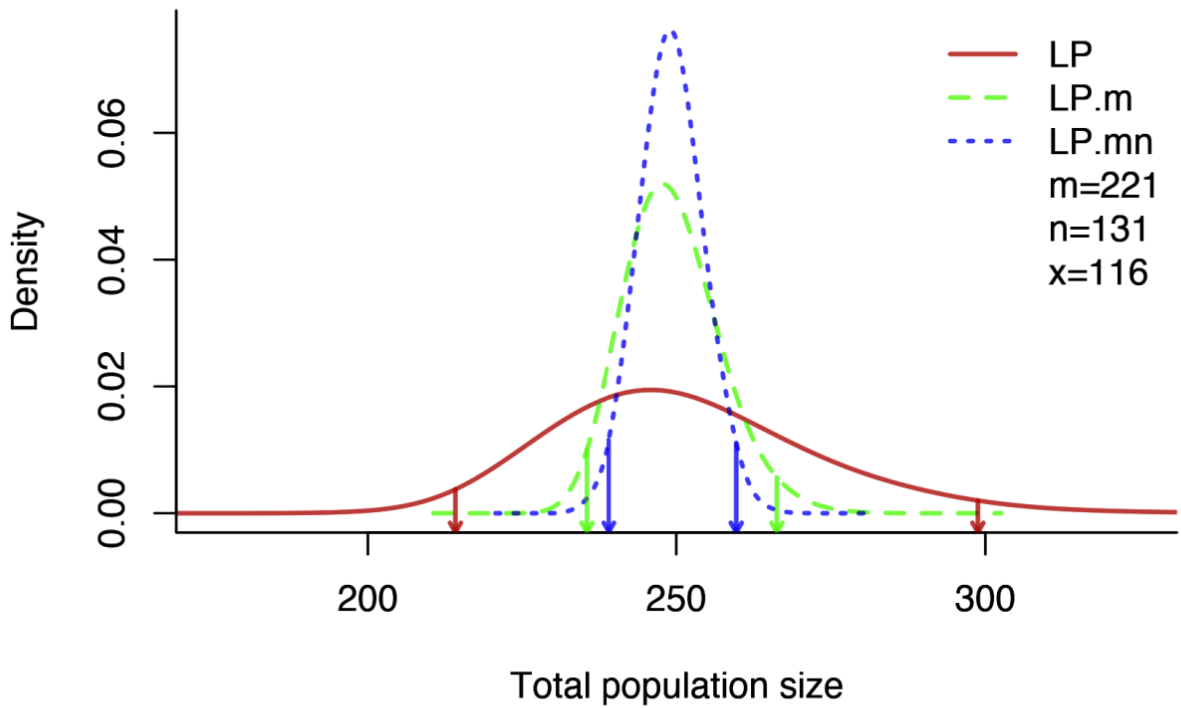

Figure Box 2.1: Sampling distributions and 95% confidence intervals of total population size estimates for three levels of conditioning in Lincoln Peterson estimates. The ML estimate for all three models is 249. The confidence 2.5% and 97.5% limits are indicated by the vertical lines dropped from each curve to the x-axis. The intervals become increasingly shorter as the models (hypothetical experiments) become more constrained. Here, as is generally but not universally the case, the intervals are completely nested.

__________________________End of Box 2 __________________________________

## 2.2 Interpreting evidential uncertainty

Generally, desirable properties in confidence intervals are proper coverage and given proper coverage, shortness of length (Cassella and Berger, 2002). A confidence interval can either cover the target or it can miss it. If the interval fails to cover the target, it can either be entirely above the target (miss high) or entirely below it (miss low) (see Figure 1). It is often, but not always, considered desirable if intervals that miss the target value are distributed equally above and below it. Evidence is one of the cases where an equal distribution of non-coverage is undesirable. In this context missing high is superior to missing low. Both types of intervals misrepresent the confidence one should have in the evidence, but the high miss is at least always indicating a correct assessment while a low miss could be supporting an incorrect assessment. Of course, this is assuming the expected evidence is positive, as in Figure 1, if the expected evidence were negative, the desirability of missing high and low would be reversed. Really, we mean that it is better for the interval to miss its target distally from 0 than to miss proximally to 0. However, in this simulation study the evidential comparisons are arranged so the reference model is always the better model as to keep the language of missing high and low less confusing.

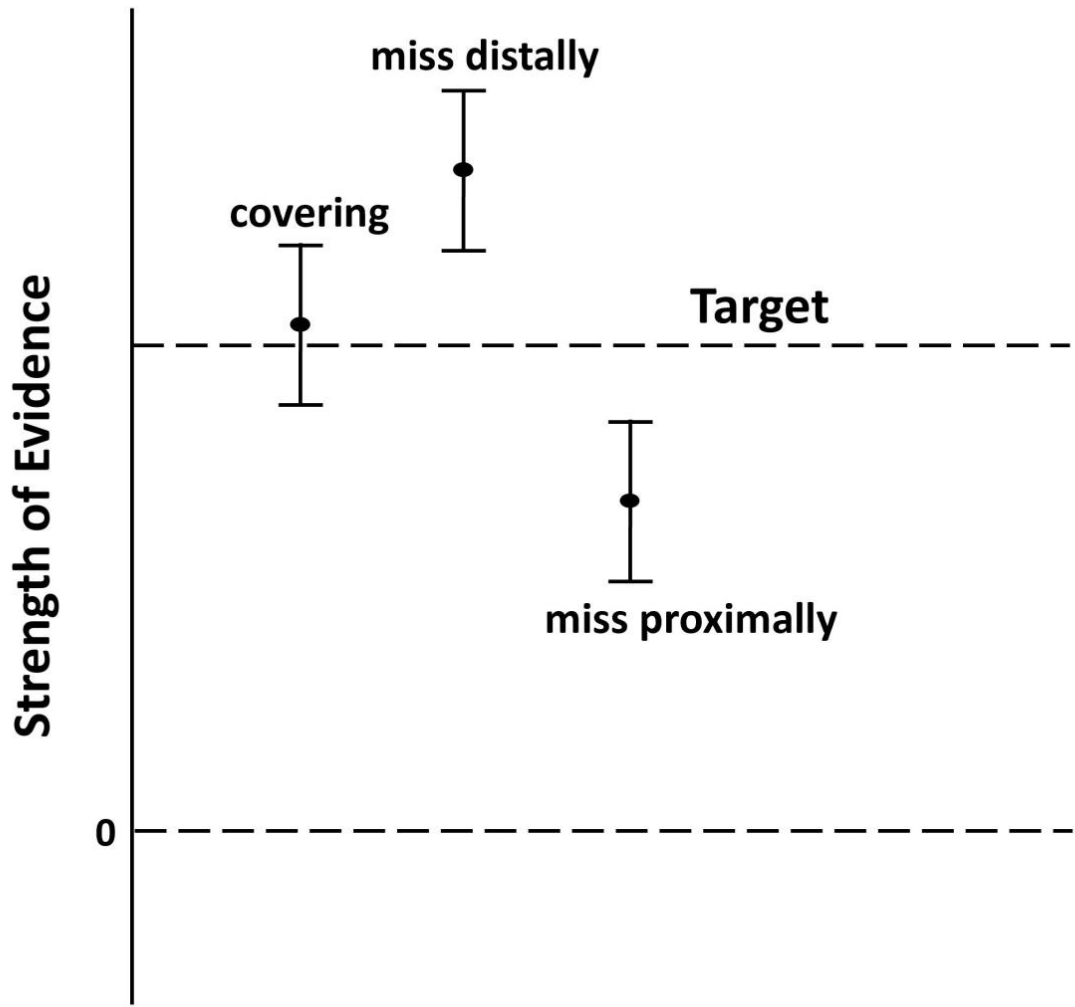

Figure 1: Hypothetical coverage of confidence intervals for evidence. The strength of evidence is the value of an evidence function relating two models and a data set. Typical evidence functions are LLR or the difference of information criterion values, $\Delta IC$s . In our worked example (section 3) we use the Schwarz information criterion. $\Delta SIC_{ra}$ values greater than 0 indicate support in the data for the reference model relative to the alternative. These values are indicated by dots in the figure. The vertical bars indicate confidence intervals for the strength of evidence. The target for a confidence interval on the strength of evidence is a penalized scaled divergence difference (see section 4.1), loosely this is the expected evidence. By design, a perfect confidence interval, at say the 95% confidence level, will fail to cover its target 5% of the time.  If a confidence interval that misses its target is entirely more distant from 0 than is its target, we say that it misses distally, otherwise we say that it misses proximally. We will also speak of the bound of a confidence interval for evidence that is closest to 0 as the proximal bound.

The categories of evidence introduced in Box 1 suggest useful ways to apply confidence intervals for strength of evidence to scientific inference. Scientifically, the paramount question is the evidence veridical (i.e. in agreement with fact) or is it misleading? The intervals we propose estimating can give us confidence in our answer. We propose that if the proximal bound of this confidence interval is distal to $k_S$ that it be considered "very secure". If the proximal bound falls between $k_S$ and $k_p$ then the evidence should be considered "secure". Finally, if the proximal bound is proximal to $k_p$ or the interval overlaps 0 the evidence is "insecure".

These three levels of strength of evidence and two levels of security of evidence create 6 heuristic categories

1. Strong and very secure (SV): The point estimate of evidence (e.g. $\Delta$SIC )is strong and the lower bound of uncertainty indicates that have confidence that the true evidence is also strong.
2. Strong and secure (SS): The point estimate of evidence is strong, and we are confident that the true target is at least prognostic. There is very little chance that this evidence is misleading.
3. Strong but insecure (SI): The point estimate of evidence is strong, but we cannot be confident that the target is not weak.
4. Prognostic and secure (PS): The point estimate of evidence is prognostic, and we can be confident that the target is at least prognostic.
5. Prognostic but insecure (PI): The point estimate of evidence is prognostic, but we are not confident that the target is not weak.
6. Weak and insecure (WI): The point estimate of evidence is weak and thus by definition, we are not confident that the target is not weak.

As sample sizes increases, majority of the sampling distribution lies above the strong evidence threshold and probability of obtaining evidence that is not SS diminishes to 0 (Dennis et al. 2019). There is, of course, the pathological case where two models are equally divergent from the true generating process. Were this curiosity ever to occur, then each model would be strongly and securely selected with probability 0.5. It is arguable that, even in such a situation, no error has occurred, as in each case a model closest to the generating process has been selected. Substantial discussion on interpreting statistical evidence when augmented with confidence intervals is given in Box 3.

---

Box 3 Interpreting Evidence using Confidence Intervals

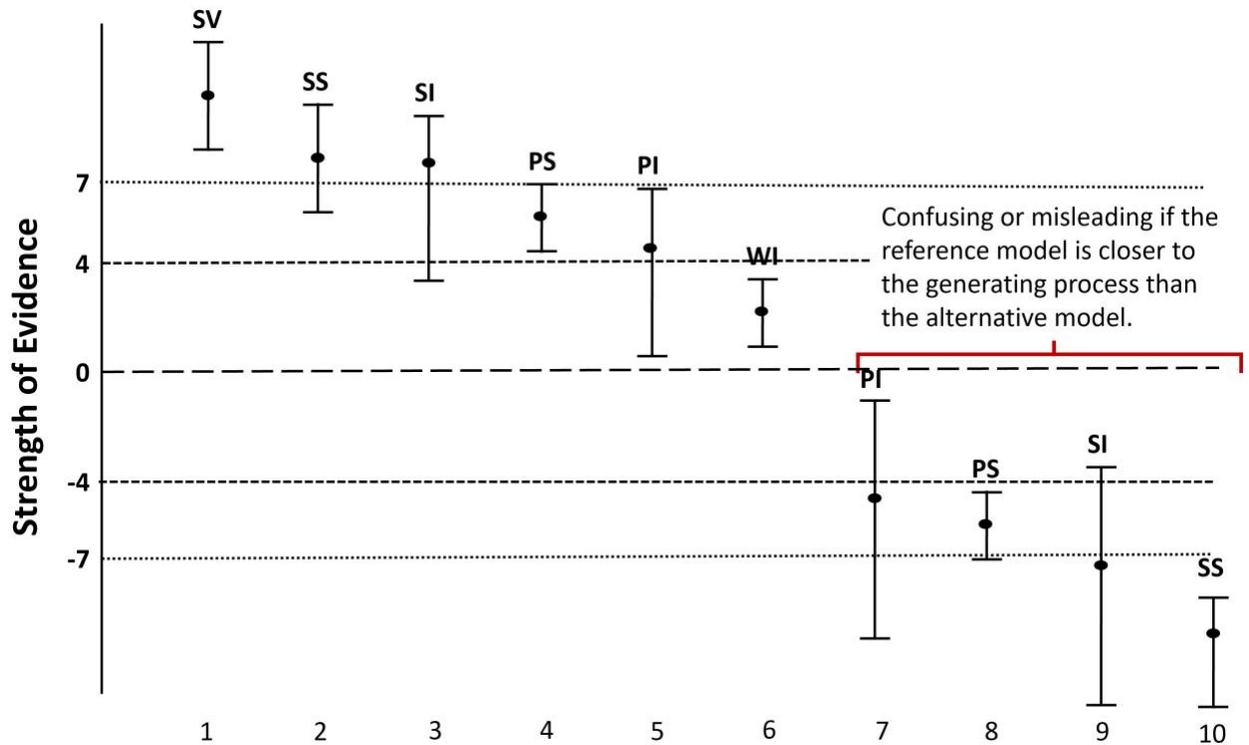

Figure Box 3.1 depicts some hypothetical confidence intervals for the strength of evidence.

In interval 1, the observed evidence (e.g. $\Delta SIC$), indicated by the filled oval, is strong and the lower bound for the confidence interval is above the strong evidence threshold. This evidence is designated strong and very secure (SV)—the reference model is strongly supported as being closer to the generating process than the alternative and there is almost no chance that sampling variation would upset this identification. In this case, the researcher may reasonably conclude that no further work is needed regarding model identification in this particular model contrast. Possibly, further work may be indicated to improve parameter estimate precision in the identified better model.

In interval 2, the observed evidence is above the strong evidence threshold, and the proximal bound is greater than the prognostic evidence threshold. We call this situation 'strong but secure' (SS). This implies that the reference model is strongly supported, and it is unlikely (but plausible) that this is due to sampling variation. Cautious but optimistic interpretation is indicated, and if possible, more data should be collected to confirm the conclusions.

In interval 3, the observed evidence is above the strong evidence threshold, but the proximal bound is less than the prognostic evidence threshold. We call this situation 'strong but insecure' (SI). This implies that while the reference model is strongly supported, it is uncertain due to sampling variation. Very cautious interpretation is indicated, and if possible, more data should be collected to confirm the conclusions.

In interval 4, the observed evidence is less the strong evidence threshold, and the proximal bound is greater than the prognostic evidence threshold. We call this situation 'prognostic but secure' (PS). This implies that while the reference model has only moderate support, it is unlikely that this is due to sampling variation. In this case, the distal bound is less than the strong evidence threshold. It is likely that both models explain the data nearly equally well, but with a slight edge to the favored model.

In interval 5, the observed evidence is less the strong evidence threshold, and the proximal bound is less than the prognostic evidence threshold. We call this situation 'prognostic but insecure' (PI). This implies that the reference model has only moderate support and even this may be due to sampling variation. The primary implication is that more data is needed either within the context of the current experiment or by combining these results with the results of other experiments.

In interval 6 the evidence is weak and insecure (WI). The models are not differentiated by the data. The researcher should collect more data in order to identify the models. The researcher should of course recognize that not all data is equally informative and seek data that will distinguish the two models (e.g. Cooper et al., 2008). Another choice that could be made, particularly if large amounts of data have already been collected, is to decide that both models are adequate for the intended purposes (Lindsay, 2004; Markatou and Sofikitou, 2019)

Intervals 7, 8, 9, and 10 are reflections of intervals 5, 4, 3, and 2, only in this case they are misleading. The designation C stands for confusing evidence, which is prognostic evidence for the wrong model. The designation M stands for misleading evidence, which is strong evidence for the wrong model.

Interval 10 is a researcher's worst case. The evidence is strong, secure and misleading. The researcher should try to avoid this situation both by experimental design (large sample size, treatments or observations that strongly differentiate between the models) and by analytic design (higher strong and marginal evidence thresholds).

In practice, we do not know if the evidence is misleading or not. For this discussion, we consider 'negative' evidence as misleading but in fact, it only indicates that it supports the alternative model— unless one knows the location of the generating process (see Ponciano and Taper 2019). Simulations (section 5.2 and Taper et al., 2019) show that for global evidence strong but secure misleading evidence occurs very rarely—regardless of whether the model set is correctly specified or misspecified. For local evidence if the model set is correctly specified, secure misleading evidence is exceedingly rare. However, under model misspecification secure misleading evidence occurs more frequently, although it is still not common. We present an explicit comparison of global and local inference under correctly specified and misspecified models in section 5.2.

___________________________End of Box 3 ________________________________________

## 3    Example: Uncertainty in an SEM analysis of post-fire recovery of plant diversity

To probe the effectiveness of bootstrapping evidence in realistically complex problems, we revisit the classic analysis of Grace and Keeley (2006). These authors used structural equation modeling to study the impact of landscape, environment, and community factors on the recovery after fire of shrubland plant diversity.

A recent article on developing causal models (Grace and Irvine, 2020) revisits the 2006 study and takes a more moderate stance: "Subsequent SEM studies (Keeley et al. 2008) have enhanced our confidence in the general inferences drawn from the original study. That said, we would not claim that all our parameter values are unbiased causal estimates without further evidence to support such inferences." We believe that had Grace and Keeley had the tools for estimating the two kinds of evidential uncertainties we have developed here a much more nuanced understanding could have been gained—even from the original data—as to which paths were likely to be supported by future work and which were potentially non-replicable.

## 3.1   Example choice

There are reasons why SEM is growing in influence in environmental informatics, ecology and evolution.  First, SEM allows for legitimate causal inference in situations both in observational studies (Grace, 2008; Bollen and Pearl, 2013, Grace and Irvine, 2020) and where experimental manipulation has been performed (Grace et al., 2009; Breitsohl, 2019).  In fact, path analysis, the precursor to SEM, was first developed by Sewall Wright (1934) to expose causal effects to statistical inference.  Second, because it is designed for estimation of a network of causal effects, SEM is well suited for analyses of the complex patterns of influence often found in environmental science, ecology and evolution (e.g. Grace and Pugesik, 1997).   Third, SEM recognizes that many observables may be recorded with measurement error. The ability to incorporate measurement error in an analysis eliminates an important source of bias that has plagued environmental science, ecology and evolution (Taper and Marquet, 1996; Cheng and Van Ness, 1999). Implicit in the incorporation of measurement error is the ability to consider latent variables (i.e. unobserved, and potentially unobservable variables).  (Grace and Bollen, 2008; Grace et al., 2010).  Fourth, causal paths and latent variables allow linking scientific theory and statistical analysis in a particularly perspicuous fashion (Grace and Bollen, 2008, Grace et al. 2010, Laughlin and Grace, 2019). Because of these beneficial features, SEM is being utilized in growing number of applications in environmental informatics, ecology and evolution. The explosive growth of SEM in ecology is documented in Laughlin and Grace (2019).

Despite its many advantages for scientific thinking, SEM does present some inferential difficulties (Tomarken and Waller, 2003).  Information can flow between variables by multiple pathways.  As a consequence, the fit of alternative models and therefore the evidence between them can vary considerably with small changes in the configurations of the data.  This uncertainty in evidence needs to be quantified.

A final reason for the choice of the Grace and Keeley example is the excellence of the original study. The observations were collected under the direction of Jon Keeley, while the analysis was conducted by James Grace.  Jon Keeley is a very experienced empirical ecologist, while Grace has been a leading proponent of the application of SEM to ecological systems.  Both are scientists of great distinction.  We do not seek to cavil at pedestrian research but look to see what bootstrapping of evidence can add to a well done scientific analysis.

## 3.2   Example description

Grace and Keeley (2006) and Keeley et al. (2005) describe the data collection in detail.  In brief, 90 sites in southern California were surveyed for 5 years following wildfire. Seven variables were observed indicating 7 latent variables (see Table 1).  Variables were transformed to generate approximate linear homoscedastic relationships.

Table 1: Descriptions of variables from Grace and Keeley 2006

| Observed variable G&K name | G&K Data file name | Latent variable G&K name | Single character abbrev. TLPD&J | Measurement error assumed |
|---|---|---|---|---|
| Distance from coast | Distance | Landscape Position | L | No |
| Age | Age | Stand Age | A | No |
| Community Heterogeneity | Hetero | Heterogeneity | H | Yes |
| Abiotic optimum | Abiotic | Local abiotic conditions | C | No |
| Fire index 1 | Firesev | Fire severity | F | Yes |
| Species/plot | Rich | Richness | R | No |
| Total cover | Cover | Plant cover | P | No |

### 3.3    Model Naming Conventions

We will use a model naming convention that indicates latent variable regression structure. The single character abbreviation for a variable will be followed by "." and then by the abbreviations for the variables it is regressed on. Regressions with different response variables will be separated by "_".

If a latent is isolated, that is it is neither a response nor a predictor in any regression in the model, its character would be entered in the model name but not followed by a ".". We don't consider any such models, because we are picking up the Grace and Keeley reanalysis mid-stream, after they eliminated a variable called "Community Type" from their analysis. Alphabetical order will be imposed so that a path model uniquely determines a name. Thus the Grace and Keeley best model can be named: "A.L_C.L_F.A_H.L_P.F_R.CHLP" (see Figure 2 and Table 1).

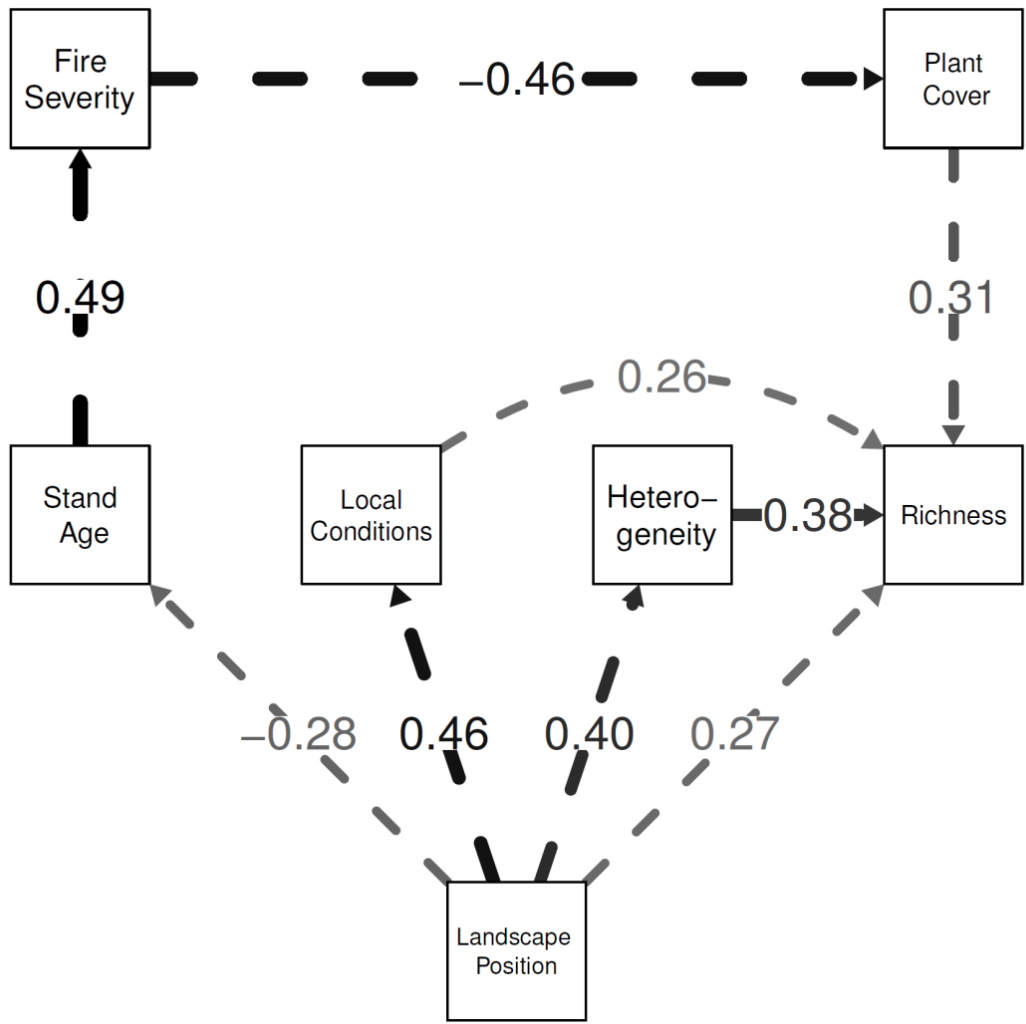

Figure 2. The final, simplified model explaining plant diversity redrawn from Grace and Keeley (2006). Arrows indicate causal influences. The standardized coefficients are indicated by path labels and widths. Weak paths with coeficients of magnitude less than 0.30 are shown in gray.

## 3.4   Example Reanalysis

Dr. Grace kindly provided the original data set and his original code (written using R package lavaan). In our reanalysis we use the R package lava (version 1.6.7). The estimates of the standardized coefficients from the two packages agree to at least the 5 decimal places reported by lava. Grace and Keeley determine their best model based on several factors including theoretical background, chi-square model adequacy tests, generalized likelihood ratio tests between nested models, and inspection of deviations between observed and model implied covariances. Grace and Keeley note the consistency of their model identification with identification based on information criterion.

The strong theoretical relationship between $\Delta IC$s, the difference of information criterion values, and the likelihood ratio test statistic has been noted before (e.g. Burnham and Anderson, 2002, Lele and Taper, 2011; Taper and Ponciano, 2016). What differs between the approaches are the assumptions and warrants that tie the statistics to scientific inference. These differences can lead to substantive differences in inference from the same data and essentially the same statistic. With a NP test your inference is a categorical accept or reject if your p-value is 0.051, just the wrong side of alpha of 0.05 your reject. If you have a $\Delta IC$ of 6.9, you don't reject it instead you give a more elaborate discussion: "Well the evidence doesn't quite reach our arbitrary strong evidence threshold, but it is very strong prognostic evidence." We will return to this in the discussion. Here we focus on the impact of uncertainty in evidence for one model over another given the data on reasonable scientific inference.

### 3.4.1 Models considered

Statistical evidence, at least defining the term in the Royall (1997), Lele (2004), Taper and Ponciano (2016) and Brittan and Bandyopadhay (2019) tradition, is not unary, but binary: It measures the support (Edwards, 1992) for one model over another model that is given by data. The models we compare are listed in Table 2.

The first model is the Grace and Keeley best model (GKBM). The next 9 models are deletion models that each differ from the best model by the absence of a single path. These models are listed in order (strongest to weakest) of the strength of the effect in the best model (as measured by the coefficient z-statistic). Comparison of each of these models with the GKBM will probe the question of whether the deleted path belongs in "best model". The last 5 models are addition models that each differ from the GKBM by the presence of 1 or 2 paths. Comparison of each of the addition models with the GKBM probes the question of whether that/those paths should be included in a "best model".

Table 2: Models compared in our reanalysis of the Grace and Keeley 2006 structural equation analysis of diversity recovery after fire. The left-hand column gives the model's full name, which indicates the complete path structure. The right-hand column describes how the model relates to the Grace and Keeley best model.

| Full name | Description |
|---|---|
| A.L_C.L_F.A_H.L_P.F_R.CHLP | GKBM (G & K best model) |
| A.L_C.L_H.L_P.F_R.CHLP | GK–M – C~L |
| A.L_F.A_H.L_P.F_R.CHLP | GK–M – F~A |
| A.L_C.L_F.A_H.L_R.CHLP | GK–M – P~F |
| A.L_C.L_F.A_P.F_R.CHLP | GK–M – H~L |
| A.L_C.L_F.A_H.L_P.F_R.CLP | GK–M – R~H |
| A.L_C.L_F.A_H.L_P.F_R.CHL | GK–M – R~P |
| C.L_F.A_H.L_P.F_R.CHLP | GK–M – A~L |
| A.L_C.L_F.A_H.L_P.F_R.HLP | GK–M – R~C |
| A.L_C.L_F.A_H.L_P.F_R.CHP | GK–M – R~L. |
| A.L_C.L_F.A_H.L_P.F_R.CFHLP | GKBM + R~F. Clarifies G&K question 4 |
| A.L_C.L_F.A_H.L_P.AF_R.CHLP | GKBM + P~A. Clarifies G&K question 7 |
| A.L_C.L_F.A_H.L_P.F_R.ACHLP | GKBM + R~A. G&K Model D |
| A.L_C.L_F.A_H.L_P.FL_R.CHLP | GKBM + P~L. Added because of covariance residuals |
| A.L_C.L_F.A_H.L_P.AFL_R.CHLP | GKBM + P~AL. Added because of covariance residuals. |

### 3.4.2 Example reanalysis results

The results of our reanalysis are presented in Figures 3, which plots the evidence ($\Delta$SIC) and its uncertainty for the GKBM relative to each of the deletion models, and Figure 4, which shows GKBM evidence and uncertainty relative to the addition models.

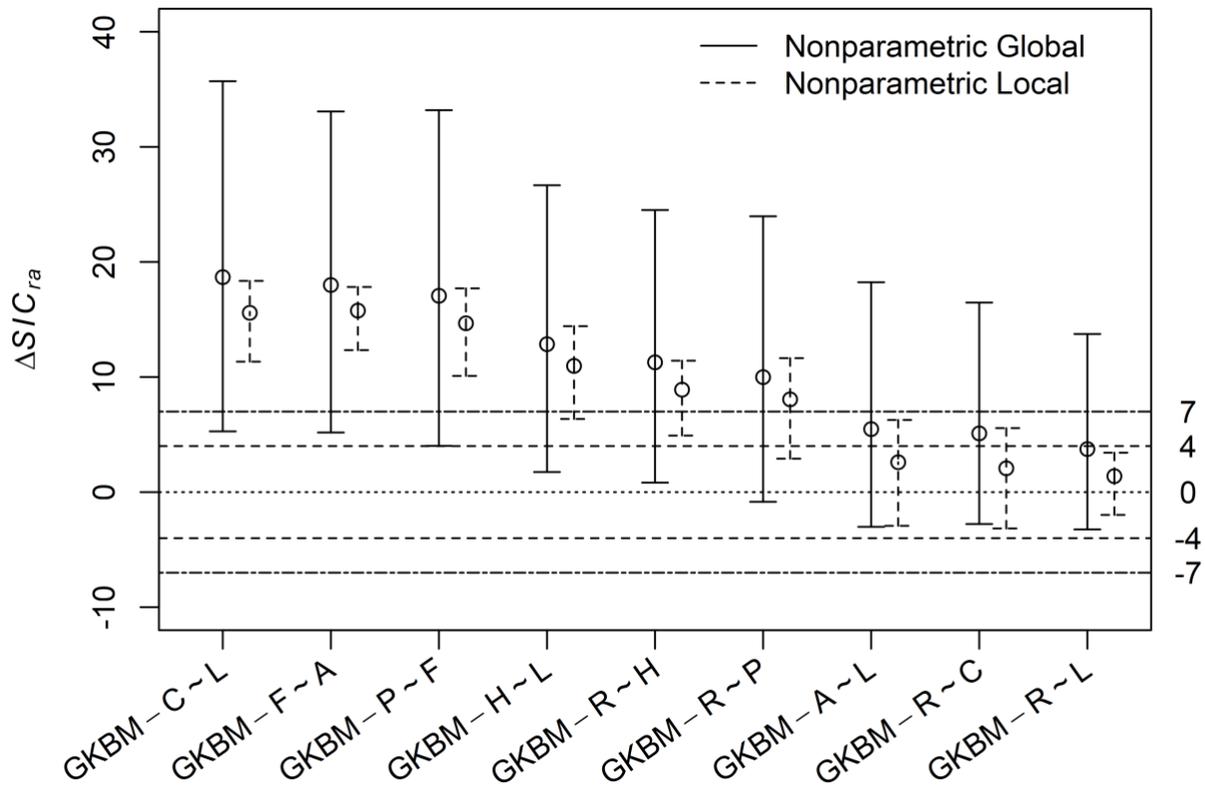

Figure 3: Evidential uncertainty intervals comparing the Grace and Keeley best model with 9 models, each that deletes one of the paths in the GKBM. For each model comparison, the open circle indicates the observed evidence, the solid error bar indicates the global uncertainty, the dashed error bars show the local uncertainty. These are approximate 90% confidence intervals based on 4000 non-parametric bootstraps. The strong evidence thresholds are indicated by dot-dash horizontal limit lines at 7 and -7, while the prognostic evidence thresholds are indicated at dashed limit lines at 4 and -4. Positive values of the $\Delta SIC_{ra}$ indicate evidence for the GKBM, as the reference model, relative to the alternative model, while negative values indicate evidence for the alternative model relative to the GKBM. The separatrix between these two regions is the dotted horizontal limit at 0.

The first three model comparisons are rock solid. They all have strong and secure global evidence and strong and very secure local evidence. Not only does this data set strongly favor including these three paths, but replication of the experiment—in the same environment—will almost always reach the same conclusion.

The next two comparisons (GKBM – H~L and GKBM – R~H) both have strong and secure local evidence for including their paths, but globally, they are insecure. We have good reason to believe that these paths represent real causal effects, but need to advise researchers seeking to replicate this experiment to increase sample size to avoid equivocal results.

Then a comparison (GKBM – R~P) with evidence, both global and local, that is strong but insecure. Here the global interval crosses the 0 line. Researchers should consider the possibility that the path may be weaker than estimated or may be non-existent.

The next two comparisons have barely prognostic evidence for their paths, but are insecure both globally and locally, with intervals that substantially overlap the line separating evidence for one model versus evidence for the other. The final comparison has positive but weak evidence for inclusion of the path. It is by definition insecure. The local evidence interval falls entirely between the two prognostic evidence thresholds. There is evidence for the path, but it is just a bit more than a toss-up.

Whether or not the last 3 paths are included in model is a judgement call for the reporting researchers based on the costs both practical and intellectual of including false paths or omitting true paths. For these deletion paths, a nudge might be given towards including them because the evidence favors the more complex model despite the SIC evidence function being used having a slight bias at small sample size towards compact models.

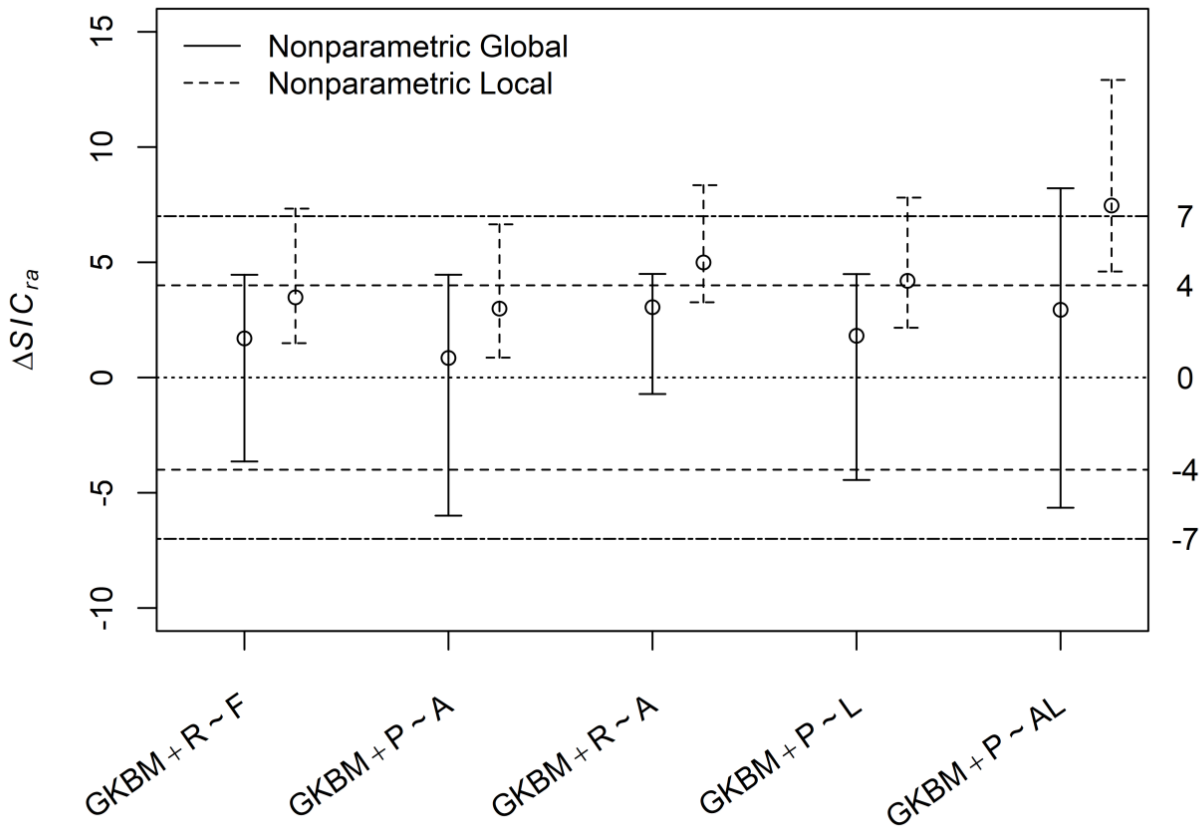

Figure 4: Evidential uncertainty intervals comparing the Grace and Keeley best model with 5 models, each that adds one or two paths to the GKBM.

All 5 addition models have global evidence that is weak and insecure but that leans towards the more compact GKBM. However, all the global intervals overlap the separatrix at 0, and three of the intervals even overlap the marginal evidence thresholds for including the paths. The local evidence shifts slightly further towards the GKBM.

At this sample size, there is no compelling statistical reason to include any of the addition paths in the "best model", but there is also no compelling statistical reason not to. The slight tilt towards the GKBM may represent nothing more that the SIC bias towards compact models. It is very hard statistically to distinguish between the true absence of a path and the presence of a weak path. It would take a sample size of more than 1000 for there to be an expectation of global strong and secure SIC evidence for the absence of a path even if it was truly absent. On the other hand, because the coefficient of variation of local evidence declines at a much faster rate than that of global evidence ($n^{-1}$ versus $n^{-1/2}$) even a modest increase in sample size may allow local identification of weak effects. In the case of the Grace and Keeley example the breadth of the conditional intervals indicates that the sample size is marginal in a statistical sense—despite the Herculean effort represented.

Models are single entities, but they are entities built from components. In our experience, a great deal of insight into how components function in models can be found by estimating the evidence for a model including the component relative to the same model without that component. In all 14 model comparisons, the weight of evidence tilts towards the GKBM. We agree with Grace and Keeley that A.L_C.L_F.A_H.L_P.F_R.CHLP is the "best model" (at least out of those considered) to describe the structural relationships in this data set. Grace and Keeley chose in 2006 to interpret the empirical results of their study narrowly. "Ultimately, results and interpretations presented in this paper are based on the model judged to be the best representation of the data" (Grace and Keeley, 2006). Here we do disagree with Grace and Keeley. Our analysis has shown that even within a small list of a priori models, drawn from their own back-ground theory, there are multiple plausible models whose interpretation should be considered. To interpret only a single best model is like choosing to use only a parameter point estimate without considering its uncertainty. It is simple, but over-confidence can be generated.

## 4    Mathematical Development

In this section, we develop the statistical justification and estimation algorithms for the confidence intervals for evidence that we use in this paper. A reader satisfied with a simulation based justification could skip to section 5, at least on first reading.

Different statistical divergences could be used to construct model adequacy measures and thus evidence functions (see Lele 2004, Markatou and Sofikitou, 2019). Each will have its own properties, and each could be useful in different circumstances. In this paper we focus on the Kullback-Leibler divergence (KLD) as it leads to the information criteria, evidence functions already in common use. The treatment of uncertainty for other divergences and evidence functions should parallel that for the KLD. The mathematical notation, definitions, and assumptions used in our treatment are given in Box 4.

---

Box 4: Mathematical notations, definitions, and assumptions

1. Data are assumed to be suitable for non-parametric bootstrapping. For this paper we further assume that the data are independently and identically distributed (i.i.d.).

2. Probability density (or mass) function representing the true generating mechanism: $g(.)$. Its cumulative distribution function (cdf) is denoted as $F_g(\cdot)$

3. Observed data: $\underline{x} = (x_1, x_2 ..., x_n)$, where $n$ denotes the sample size.

4. Random variables: $\underline{X} = (X_1, X_2, ..., X_n)$.

5. The pdfs (or pmfs) for reference ® and alternative ($A$) models are denoted by $m_R(.)$ and $m_A(.)$ respectively. For example, $m_R$ is $N(\mu = 5, \sigma = 1)$.

6. Reference and alternative model spaces are denoted by $M_R$ and $M_A$ where each is a collection of models. For example $M_R = \left\{ N(\mu, \sigma) \text{ with } \mu \text{ in } (-\infty, \infty) \text{ and } \sigma \text{ in } (0, \infty) \right\}$.

7. $F_g^{(n)}(t; \underline{X}) = \frac{1}{n} \sum_{i=1}^{n} I(X_i \le t)$ is the empirical estimator of the cdf of $g(.)$ Where $I(A)$ is the indicator function for event $A$. Denote the corresponding smoothed density as $g_{n,\underline{X}}$.

8. $\hat{F}_g^{(n)}(t; \underline{x}) = \frac{1}{n} \sum_{i=1}^{n} I(x_i \le t)$, the empirical estimate of the cdf of $g(.)$. Denote the corresponding smoothed density as $\hat{g}_{n,\underline{x}}$.

9. The KLD between two specified continuous models, where the reference model is $m_1$ is $K(m_1, m_2) = \int \left( \log(m_1(x)) - \log(m_2(x)) \right) m_1(x) dx$. In general, for any two models (discrete, continuous, or piecewise continuous) we write $K(m_1, m_2) = \int \left( \log(m_1(x)) - \log(m_2(x)) \right) dF_{m_1}(x)$

10. The KLD orthogonal projection onto a model space $M$ from of a specified model $s(.) \notin M$, is $K(s(.), M) = \min_{m \in M} K(s(.), m)$ (see Figure 3 in Ponciano and Taper, 2019).

11. Denote the model in model space $M$ at which the above minimum is attained as $m_s^*(.)$. If $s(.) \in M \Rightarrow m_s^*(.) \equiv s(.)$

12. The log-likelihood function for the observed data under $g(.)$ is $l_g(\underline{x}) = \sum_{i=1}^{n} \log(g(x_i))$

   The log-likelihood function for the observed data under a model $m(.)$ is

   $l_m(\underline{x}) = \sum_{i=1}^{n} \log(m(x_i))$. $\hat{m}(.)$ is the model with parameter values that maximizes $l_m(\underline{x})$.

13. If one partitions the parameter vector, $\theta$, for a model, $m$, into a vector of parameters of interest, $\gamma$, and a vector of nuisance prameters, $\lambda$, then the profile log-likelihood for $\gamma$ can be written as $l_{p,m}(\gamma; \underline{x}) = \max_{\lambda} l_m(\underline{x}; \gamma, \lambda)$.

14. The KLD estimator of the divergence of a model, $m$, from the generating process, $g$:
$$K(g_{n,\underline{X}}, m; \underline{X}) = \int \log(g_{n,\underline{X}}(t)) dF_g^{(n)}(t; \underline{X}) - \int \log(m(t)) dF_g^{(n)}(t; \underline{X})$$
$$= S_{g_{n,\underline{X}}, g_{n,\underline{X}}} - S_{g_{n,\underline{X}}, m},$$
where $S_{g_{n,\underline{x}} g_{n,\underline{x}}}$ is the neg-self-entropy of the empirical distribution.

15. The KLD estimate of the divergence of a model, *m*, from the generating process, *g*:

$$K(\hat{g}_{n,\underline{x}}, m; \underline{x}) = \int \log\left(\hat{g}_{n,\underline{x}}(t)\right) dF_g^{(n)}\left(t; \underline{x}\right) - \int \log\left(m(t)\right) dF_g^{(n)}\left(t; \underline{x}\right)$$

$$= S_{\hat{g}_{n,\underline{x}}, \hat{g}_{n,\underline{x}}} - S_{\hat{g}_{n,\underline{x}}, m}$$

where $S_{\hat{g}_{n,\underline{x}} \hat{g}_{n,\underline{x}}}$ is the neg-self-entropy of the empirical distribution.

16. The KLD projection estimator of the divergence of a model space, *M*, from the generating process, *g*: $K(g_{n,\underline{X}}, M; \underline{X}) = S_{g_{n,\underline{X}}, g_{n,\underline{X}}} - S_{g_{n,\underline{X}}, m^*_{g_{n,\underline{X}}}}$

17. The KLD projection estimate of the divergence of a model space, *M*, from the generating process, *g*: $K(\hat{g}_{n,\underline{x}}, M; \underline{x}) = S_{\hat{g}_{n,\underline{x}}, \hat{g}_{n,\underline{x}}} - S_{\hat{g}_{n,\underline{x}}, m^*_{\hat{g}_{n,\underline{x}}}}$

18. One estimate for $K(\hat{g}_{n,\underline{x}}, M; \underline{x})$ is $S_{\hat{g}_{n,\underline{x}} \hat{g}_{n,\underline{x}}} - l_{\hat{m}(\underline{x})}$. Bias correction for this estimate is the goal of information criteria.

19. The global penalized scaled divergence difference target:

$$\Delta D_{Pn}(g, M_R, M_A) = 2n\{K(g, M_A; \underline{X}) - K(g, M_R; \underline{X})\} + c_n(p_A - p_R)$$. Note that for fully specified model comparisons, the penalty term is 0, and

$$\Delta D_{Pn}(g, m_R, m_A; \underline{X}) = 2n\{K(g, m_A; \underline{X}) - K(g, m_R; \underline{X})\}$$

20. The local penalized scaled divergence difference target,

$$\Delta d_{Pn}(g, M_R, M_A, \underline{x}) = 2n\{K(g, m_A^*, \underline{x}) - K(g, m_R^*, \underline{x})\} + c_n(p_A - p_R)$$, For fully specified models $\Delta d_{Pn}(g, m_R, m_A, \underline{x}) = 2n\{K(g, m_A, \underline{x}) - K(g, m_R, \underline{x})\}$

21. The global penalized divergence difference estimator,

$$\Delta D_{Pn}(g_{n,\underline{X}}, M_R, M_A, \underline{X}) = E_{g_{n,\underline{X}}}\left(2n\{K(g_{n,\underline{X}}, m_{A,\underline{X}}^*, \underline{X}) - K(g_{n,\underline{X}}, m_{R,\underline{X}}^*, \underline{X})\} + c_n(p_A - p_R)\right)$$, For fully specified models $\Delta D_{Pn}(g_{n,\underline{X}}, m_R, m_A, \underline{X}) = E_{g_{n,\underline{X}}}\left(2n\{K(g_{n,\underline{X}}, m_A, \underline{X}) - K(g_{n,\underline{X}}, m_R, \underline{X})\}\right)$

22. The local penalized divergence difference estimator,

$$\Delta d_{Pn}(g_{n,\underline{X}}, M_R, M_A, \underline{x}) = E_{g_{n,\underline{x}}}\left(2n\{K(g_{n,\underline{X}}, m_{A,\underline{X}}^*, \underline{x}) - K(g_{n,\underline{X}}, m_{R,\underline{X}}^*, \underline{x})\} + c_n(p_A - p_R)\right)$$, For fully specified models $\Delta d_{Pn}(g_{n,\underline{X}}, m_R, m_A, \underline{x}) = E_{g_{n,\underline{x}}}\left(2n\{K(g_{n,\underline{X}}, m_A, \underline{x}) - K(g_{n,\underline{X}}, m_R, \underline{x})\}\right)$

23. The global evidence estimate,

$$Ev_G\left(M_R, M_A; \hat{g}_{n,\underline{x}}, \underline{x}\right) = E_{\hat{g}_{n,\underline{x}}}\left(2n\{K(\hat{g}_{n,\underline{x}}, \hat{m}_{A,\underline{X}}, \underline{X}) - K(\hat{g}_{n,\underline{x}}, \hat{m}_{R,\underline{X}}, \underline{X})\} + c_n(p_A - p_R)\right)$$. Note

$$= E_{\hat{g}_{n,\underline{x}}}\left(-2\{l_{\hat{m}_{A,\underline{X}}}\left(\underline{X}\right) - l_{\hat{m}_{R,\underline{X}}}\left(\underline{X}\right)\} + c_n(p_A - p_R)\right)$$

that inside the expectation $\underline{X}$ is a random vector drawn from $\hat{g}_{n,\underline{x}}$ and that the maximum likelihood estimate, $\hat{m}$, has been substituted for $m^*$ (see definition 18). Variation in $Ev_G$ is due to variation in $\hat{m}_{A,\underline{X}}$, $\hat{m}_{R,\underline{X}}$, and $\underline{X}$. For fully specified models,

$$Ev_G\left(m_R, m_A; \hat{g}_{n,\underline{x}}\right) = E_{\hat{g}_{n,\underline{x}}}\left(2n\{K(\hat{g}_{n,\underline{x}}, m_A, \underline{X}) - K(\hat{g}_{n,\underline{x}}, m_R, \underline{X})\}\right)$$. Positive values for evidence indicate that the reference model is supported over the alternative model (see discussion Box 1).

24. The local evidence estimate,

$$Ev_L\left(M_R, M_A; \hat{g}_{n,\underline{x}}, \underline{x}\right) = E_{\hat{g}_{n,\underline{x}}}\left(2n\{K(\hat{g}_{n,\underline{x}}, \hat{m}_{A,\underline{X}}, \underline{x}) - K(\hat{g}_{n,\underline{x}}, \hat{m}_{R,\underline{X}}, \underline{x})\} + c_n(p_A - p_R)\right)$$. Note that

$$= E_{\hat{g}_{n,\underline{x}}}\left(-2\{l_{\hat{m}_{A,\underline{X}}}\left(\underline{x}\right) - l_{\hat{m}_{R,\underline{X}}}\left(\underline{x}\right)\} + c_n(p_A - p_R)\right)$$

inside the expectation $\underline{X}$ is a random vector drawn from $\hat{g}_{n,\underline{x}}$ and that the maximum likelihood estimate, $\hat{m}$ , has been substituted for $m^*$ (see definition 18). Variation in $Ev_L$ is due to variation only in $\hat{m}_{A,\underline{X}}$ and $\hat{m}_{R,\underline{X}}$. For fully specified models,

$$Ev_L\left(m_R, m_A; \hat{g}_{n,\underline{x}}, \underline{x}\right) = E_{\hat{g}_{n,\underline{x}}}\left(2n\{K(\hat{g}_{n,\underline{x}}, m_A, \underline{x}) - K(\hat{g}_{n,\underline{x}}, m_R, \underline{x})\}\right).$$ Positive values for evidence indicates that the reference model is supported over the alternative model (see discussion Box 1).

_______________________End of Box 4 _______________________________________

Commonly, either confidence or credible intervals are used to quantify uncertainty in parameter estimates. A very general method of constructing confidence intervals is hypothesis test inversion (Casella and Berger, 2002). If your test is a generalized likelihood ratio test then the set $\left\{\theta, 2\left(l_{m_{\hat{\theta}}}\left(\underline{x}\right) - l_{m_{\theta}}\left(\underline{x}\right)\right) < \chi^2_{p,(1-\alpha)}\right\}$ is an approximate $100(1-\alpha)\%$ confidence interval if $\theta$ is of dimension 1 or confidence region if $\theta$ is of dimension >1 (Pawitan 2001).

If one is interested in inference on a subset of the parameters in a multidimensional parameter vector $\theta$, one can partition the parameter vector as $\theta = [\gamma, \lambda]$, where $\gamma$ is a vector of the parameters of interest, often of dimension 1, and $\lambda$ is a vector of all the other parameters. A profile log-likelihood (for a given $\gamma$)can be calculated as $l_{p,m}(\gamma; \underline{x}) = \max_{\lambda} l_m(\underline{x}; \gamma, \lambda)$, that is by maximizing over $\lambda$. It is argued (Cox and Reid, 1987) that maximization of the profile likelihood leads to inconsistent estimators of the parameters of interest because it does not appropriately penalize for the cost of the estimation of the incidental parameters. Various bias corrections or penalty terms for the profile likelihood have been suggested (Pace and Salvan 2006).

The connection between profile likelihood and model selection becomes obvious if one considers that the parameter of interest could be nothing more than an index for the models considered. In Box 5 we use this connection to develop and justify global and local uncertainty in the evidence for one model over another (see Meeker and Escobar (1995), for a gentle introduction and Pierce and Bellio (2017) for a substantial review). We point out that these penalties for parameter estimation are similar to the penalties employed in information criteria. A general parametric bootstrap approach to calculating an approximate penalty for the profile likelihood is described in Pace and Salvan (2006).

___________________________________________________________________________

Box 5: Adjusted profile likelihood for model selection inference.

Readers can see Meeker and Escobar (1995), for a brief introduction to profile likelihood in the context of confidence interval construction and Pierce and Bellio (2017) for a substantial review of practical likelihood adjustments. A gentle introduction to model selection through information criteria can be found in Anderson (2008), with more technically robust discussions in Burnham and Anderson (2002) and Konishi and Kitagawa (2008).

A general parametric bootstrap approach to calculating an approximate penalty for the profile likelihood is described in Pace and Salvan (2006) and outlined below.

Let $M_\varphi, \varphi = 1, 2, ..., S$ denote $S$ distinct model spaces. The goal of model selection is to use the data to select the best model space. The form of the best model space is used to draw various statistical and scientific inferences about the generating mechanism.

First, we show that model selection procedure can be looked upon as a profile likelihood estimation procedure. Let $\{\underline{\theta}_1, \underline{\theta}_2, ..., \underline{\theta}_S\}$ denote the parameters for the respective model spaces $(M_1, M_2, ..., M_S)$. Denote the dimension of $\underline{\theta}_\varphi$ by $p_\varphi$.

A universal model space, that is simply a union of the model spaces, may be written as $M = \{f(x; \varphi, \underline{\theta}_\varphi), \varphi = 1, 2, ..., S\}$. In this notation, $f(x; 1, \underline{\theta}_1)$ indicates the parametric form of the probability model in the first model space, say $LogNormal(\mu, \sigma^2)$, $f(x; 2, \underline{\theta}_2)$ denotes the parametric form of the probability model in the second model space, say $Gamma(\mu, \phi)$, and so on. The parameter $\varphi$, which is a discrete parameter, is simply an index for the model space. Thus, model selection can be viewed as selecting a particular value of $\varphi$. In model selection problem, the index parameter $\varphi$ is of interest and model parameters $\underline{\theta}_\varphi$ are t'e 'inciden'al' parameters. The profile likelihood of the index parameter $\varphi$ can be written as: $l_p(\varphi, \hat{\underline{\theta}}_\varphi; \underline{x}) = \max_{\underline{\theta}_\varphi} \sum_{i=1}^{n} \log f(x_i; \varphi, \underline{\theta}_\varphi)$.

In the familiar example of the maximum likelihood estimator of the variance $\sigma^2$ in the multiple linear regression model $Y_i = \beta_0 + \beta_1 X_{1i} + \beta_2 X_{2i} + ... + \beta_p X_{pi} + \grave{o}_i$ where $\grave{o}_i \sim N(0, \sigma^2)$ independent, $\hat{\sigma}^2 = \frac{1}{n} \sum_{i=1}^{n} \left( y_i - \hat{\beta}_0 + \hat{\beta}_1 x_{1i} + \hat{\beta}_2 x_{2i} + ... + \hat{\beta}_p x_{pi} \right)^2$. This is a biased estimator and bias is pronounced when the number of covariates is large. A bias corrected profile likelihood yields the usual unbiased estimator with the divisor $(n - p - 1)$, instead of $n$. We lose $(p + 1)$ degrees of freedom because we spend some of the information in the data to estimate t'e 'nuisa'ce' parameters $(\beta_0, \beta_1, ..., \beta_p)$.

We describe the Pace-Salvan approach for the general profile likelihood case where the parameter of interest may or may not be discrete. To reflect this generality, for the description of the Pace-Salvan approach, we make a slight change in the notation. We use $\gamma$ for the parameter of interest, $\lambda$ for the incidental parameters and $h(.)$ denotes the parametric probability function presumed to be the data generating mechanism.

Let $X \sim h(., \gamma, \lambda)$. Let the parameter of interest, $\gamma$, be of dimension 1 and the nuisance parameter $\lambda$ be a vector of any dimension that does not depend on the sample size. Let $\underline{x} = (x_1, x_2, ..., x_n)$ be a random sample of size $n$ from $h(., \gamma, \lambda)$. The Log-Profile likelihood for $\gamma$ is defined as

$$l_p(\gamma; x) = \max_\lambda \sum_{i=1}^{n} \log(h(\gamma, \lambda; x_i)).$$

Model selection based on the maximum of this profile likelihood would correspond to selecting the model space that maximizes the log-likelihood but without any penalty for the number of parameters in the model. This procedure is known to lead to what is termed an inconsistent model selection procedure. The reason for the inconsistency is that this profile likelihood is a biased estimator of the expected Kullback-Leibler divergence (Akaike, 1973; see discussion in Ponciano and Taper 2019).

The inconsistency of and the bias correction used in information-based model selection bears strong similarity to the inconsistency and bias corrections in the profile likelihood estimators (e.g. Severini, 2000 or Pace and Salvan, 2006) suggested in a very different context.

Following Pace and Salvan (2006), the adjusted profile likelihood, adjusted for the effects of estimation of the nuisance parameter $\lambda$, can be computed, assuming the presumed model is

the true generating mechanism, using parametric bootstrap as follows:

1) Estimate the full parameter vector $(\hat{\gamma}, \hat{\lambda})$.

2) For each bootstrap iteration $b \in \{1, \cdots, B\}$

    a. Generate a random sample of size $n$ from $h(., \gamma, \hat{\lambda})$ denoted by $\underline{x}_b = (x_{b,1}, ..., x_{b,n})$.

    b. For these new data and for a fixed value of $\gamma$, obtain $\hat{\lambda}_b(\gamma)$ by $\underset{\lambda}{max} \sum_{i=1}^{n} \log(h(\gamma, \lambda; x_{b,i}))$.

3) Compute the simulation adjusted profile likelihood as: $l_{SA}(\gamma; \underline{x}) = \frac{1}{B} \sum_{b=1}^{B} \sum_{i=1}^{n} \log(h(\gamma, \hat{\lambda}_b(\gamma); x_i))$.

    We point out specifically that the likelihood is evaluated for the original data $\underline{x}$ but with the parameters $(\gamma, \hat{\lambda}_b(\gamma))$ that are estimated using the bootstrap data.

Pace and Salvan (2006) suggest using $l_{SA}(\gamma; \underline{x})$, instead of $l_p(\gamma; \underline{x})$ to conduct statistical inference for $\gamma$, the parameter of interest. Most importantly, they use sophisticated mathematics to show that the adjustment achieved by $l_{SA}(\gamma; \underline{x})$ is locally (conditionally, post-data, post-experiment) appropriate. Note that following Efron and Tibshirani's (1993) description of bootstrap bias correction, one may use $l_A(\gamma; \underline{x}) = 2l_p(\gamma; \underline{x}) - l_{SA}(\gamma; \underline{x})$. It follows from the results in section 3.4 of Pace and Salvan (2006) that these two versions are equivalent up to $O(n^{-1})$ and that the difference between these central estimates is small compared to the uncertainty. We use the mean of the bootstrap distribution as our central estimate to be consistent with both Pace and Salvan (2006) and Kitagawa and Konishi (2010). There is reason to believe that the median of the bootstrap distribution might have superior theoretical properties (De Blasi and Schweder, 2018), but we will pursue this in another paper.

We point out that these penalties to the profile likelihood for parameter estimation are similar to the penalties employed in information criteria. In the information theoretic literature, non-parametric bootstrap bias corrections have been developed as the extended information criterion (EIC), (Ishiguro et al., 1997; Konishi and Kitagawa, 2008; and Kitagawa and Konishi, 2010). There are two important, differences between the basic (EIC) and the Pace-Salvan adjusted profile likelihood. First, EIC uses non-parametric bootstrap whereas Pace and Salvan use parametric bootstrap. The use of non-parametric bootstrap relaxes the assumption that the parametric model is the true generating mechanism. Model misspecification is built into the EIC correction. And second, bias correction in EIC is a global (unconditional, pre-data, pre-experiment) adjustment, averaging over the variation from one experiment to other, whereas the Pace-Salvan adjustment is a local (conditional, post-data, post-experiment) adjustment that evaluates the likelihood at the observed data $\underline{x}$ but is averaged over variation of the incidental parameter estimates from one bootstrap sample to the other.

The bias correction for the EIC can be decomposed into three components: $D_1$, $D_2$, $D_3$ (Kitagawa and Konishi, 2010). One component, $D_2$, has expectation 0 and is discarded in the $EIC_2$, the variance reduced form of the EIC. The EIC bootstrap bias correction can be applied not just to the likelihood of the data, but to any functional of the data. Some algebra on equations 44 and 51 of Kitagawa and Konishi (2010) shows that . $Ev_L\left(M_R, M_A; \hat{g}_{n,\underline{x}}, \underline{x}\right) = EIC_2\left(\Delta SIC_{ra}\left(\underline{x}\right)\right) + D_1\left(\Delta SIC_{ra}\left(\underline{x}\right)\right)$. We have found numerically that $D_1\left(\Delta SIC_{ra}\left(\underline{x}\right)\right)$ is a small term that appears to have mean at or near 0, at least under the conditions that we have investigated. The SIC includes an analytic bias correction to the likelihood accounting for the number of parameters estimated. Thus, that $D_1\left(\Delta SIC_{ra}\left(\underline{x}\right)\right)$ is small in these cases does not mean that $D_1$ is always unimportant, just that we are in a region of model space where the analytic bias correction works well. Central estimates for evidence and uncertainty intervals could be based on the entire $EIC_2$. We will explore these connections elsewhere.

The Pace-Salvan adjusted profile likelihood, the EIC and the $EIC_2$ use the bootstrap distribution only to compute the bias correction factor. We stress the use of the entire bootstrap distribution to quantify uncertainty in the evidence. A non-parametric bootstrap procedure similar to the Pace-Salvan approach yields local uncertainty while a bootstrap similar to the EIC can give us global uncertainty.

_______________________End of Box 5_______________________________________

## 4.1 Divergence difference, Penalized divergence difference and Evidence functions

We start with describing precisely the quantities that we want to estimate and their estimators. Then we describe how one can obtain the sampling distribution of these estimators, either asymptotically as was done by Royall (1997, 2000) and Dennis et al. (2019) or by non-parametric bootstrap as was suggested by Taper and Lele (2011).

### 4.1.1 Fully specified competing models

Consider the case where the competing models are fully specified. In the following, we explicitly define the target quantity, its estimator (the evidence function) and the estimate (observed value of the evidence function). As has been discussed in various papers (Lele, 2004, Taper and Lele 2004, 2011, Dennis et al. 2019), the sample size scaled difference between the divergences from the true generating mechanism and the two competing hypothesized mechanisms, namely, $\Delta D_{Pn}\left(g, m_R, m_A\right) = 2n\{K\left(g, m_A\right) - K\left(g, m_R\right)\}$ is of great interest. We call this the penalized scaled divergence difference (see Box 4, definition19). This is an unknown quantity because in practice, we do not know the true generating mechanism $g(.)$.

In this formulation, because of the sample size multiplier $2n$, $\Delta D_{Pn}\left(g, m_R, m_A\right)$ converges to $\pm\infty$ or 0 as the sample size increases. We use the above formulation to be consistent with the discussion in Dennis et al. (2019) and information-based model selection criteria.

One could, alternatively, standardize the evidence so that it converges to a constant: 0 if the two models are equidistant from the true generating model, a positive number if $m_R$ is closer to $g(.)$ or a negative number if $m_A$ is closer to $g(.)$ as was done in Lele (2004). One can also use other forms of divergences such as the Hellinger divergence to quantify evidence (Lele, 2004) to make it model robust or outlier robust.

Given the data $\underline{X}$, a natural estimator of $\Delta DP_{Pn}(g, m_R, m_A)$, termed the evidence function (Lele 2004), is sample sized scaled difference of the KLD estimators (Box 4, definition 21) $2n\{K(g, m_A; \underline{X}) - K(g, m_R; \underline{X})\}$. Notice that, with the KL divergence, the unknown density $g(.)$ gets cancelled while taking the difference and does not need to be estimated explicitly. Hence this estimator reduces to the usual LLR: $Ev^{raw}(m_R, m_A; \hat{g}_{n,\underline{X}}, \underline{X}) = -2\left(l_{m_A}(\underline{X}) - l_{m_R}(\underline{X})\right)$. The estimate of the sample size scaled divergence difference, under the KLD, is:

$$Ev^{raw}(m_R, m_A; \hat{g}_{n,\underline{x}}, \underline{x}) = -2\left(l_{m_A}(\underline{x}) - l_{m_R}(\underline{x})\right).$$

In the following, we will describe the use of non-parametric bootstrap to calculate a more accurate estimate of the evidence for the reference model relative to the alternative than the raw evidence and also to quantify uncertainty in the estimated evidence as was suggested in Taper and Lele (2011). Box 6 lists an explicit algorithm for this bootstrap.

Instead of the LLR as the estimated evidence, we use the expectation (mean) of the density function of the bootstrap evidences as the estimated evidence. This could be estimated as the bootstrap average evidence. For a slight increase in accuracy, we calculate the expectation by numerically integrating over an estimated density function for the bootstrapped evidence. We use the R package kde1d (version 1.0.2, Nagler and Vatter 2019), which uses univariate local polynomial(log-quadratic) kernel density estimators. Our validation tests support the literature (Geenens and Wang, 2018) on the strength of this method. We find that confidence bounds are located more accurately with kde1d quantiles than with raw bootstrap quantiles and that estimated distributions are more accurate (in integrated squared error) than standard kernel density estimation.

We note a few important features of the bootstrapping procedure described above. When the models are fully specified the log-likelihood ratio is a U-statistics (Serfling 1984) and hence it is an unbiased estimator of the target quantity. However, if divergences other than KLD may lead to biased estimators of the target quantity. In which case, the mean of the bootstrap distribution is a bias corrected estimate of the target quantity. Also, if the models are not fully specified, it is well known that the log-likelihood ratio is a biased estimator of the target quantity (Akaike 1971). The mean of the bootstrap distribution of the log-likelihood ratio corrects for bias (Ishiguro et al. 1997).

We do not discuss the case of fully specified models any further but move on to the interesting case where parameters need to be estimated.

### 4.1.2 Competing models with unknown parameters

Next, we consider the problem of model selection where there are unknown parameters that need to be estimated. When we are dealing with model selection, the quantity of interest is scaled divergence difference penalized for the complexity of the models. We consider global penalized scaled divergence differences of the form:

$\Delta D_{Pn}(g, M_R, M_A) = 2n\{K(g, M_A) - K(g, M_R)\} + c_n(p_A - p_R)$ where $c_n$ is a function of the sample size that converges to infinity at the rate strictly between $\log(\log(n))$ and $n$ (Nishii, 1988), $p_R$ and $p_A$ are the number of unknown quantities (parameters) in the models that are estimated using the data. For example, for the Schwarz Information Criterion (SIC), $c_n = \log(n)$. This constraint guarantees that the information criterion will be a consistent criterion; that is, asymptotically it will lead to identifying the model in the model space that is closest to the true generating mechanism. We include the multiplier 2 to keep it consistent with common information criteria. We emphasize again that $\Delta D_{Pn}(g, M_R, M_A)$ is unknown in practice.

Assuming that the observations in the data are independent, identically distributed random variables, using the SIC(a.k.a. Bayesian Information Criterion or BIC) sample size correction, and using the maximum log-likelihood as an estimator of the KLD of a model to the generating process, leads to the evidence function $Ev_G(M_R, M_A; \hat{g}_{n,\underline{x}}, \underline{x}) = E_{\hat{g}_{n,\underline{x}}}\left(-2\left\{l_{\hat{m}_{A,\underline{X}}}(X_i) - l_{\hat{m}_{R,\underline{X}}}(X_i)\right\} + \log(n)(p_A - p_R)\right)$.

Where $\hat{m}_{R,\underline{X}}$ and $\hat{m}_{A,\underline{X}}$ are those models in $M_R$ and $M_A$ that are closest to $\hat{F}_g^{(n)}(.)$, the empirical CDF based on the data $\underline{X} = (X_1, X_2, ..., X_n)$, a random vector of length $n$ from $\hat{g}_{n,\underline{x}}$. Note that inside the expectation $\underline{X}$ is a random vector drawn from $\hat{g}_{n,\underline{x}}$ and that the maximum likelihood estimate, $\hat{m}$, has been substituted for $m^*$ (see definition 18). Variation in $Ev_G$ is due to variation in $\hat{m}_{A,\underline{X}}$, $\hat{m}_{R,\underline{X}}$, and $\underline{X}$. We calculate the expectation by numerically integrating over an estimated density function for the bootstrapped $\Delta SIC_{ra}$ s. We use the R package kde1d for the density estimation. Figure 5 presents a schematic of this development.

We point out that, except for the nuance of kernel density smoothing, the algorithm we describe above for $Ev_G$ is the EIC algorithm of Ishiguro et al. (1997) applied to $\Delta IC$ s rather than directly to log-likelihoods. Kitagawa and Konishi (2010) point out that the bootstrap bias correction can be applied to any functional, not just the log-likelihood. The use of the expectation of the sampling distributions of $\Delta ICs$, which already contain an analytic bias correction, adds another layer of bias correction. Accordingly, the evidence should be 3[rd] order accurate (Kitagawa and Konishi, 2010).

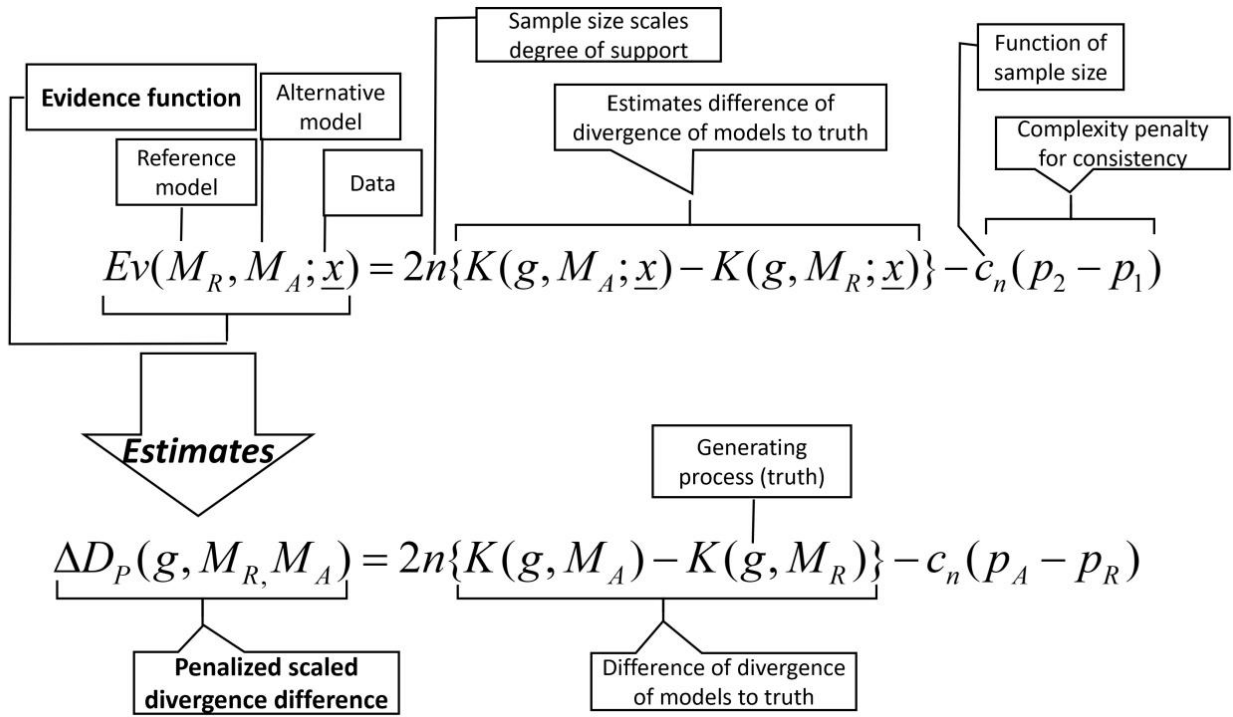

Figure 5: A schematic indicating how An evidence function relates to its target.

Similarly, the local evidence function $Ev_L\left(M_R, M_A; \hat{g}_{n,\underline{x}}, \underline{x}\right)$ is an estimate of the local penalized scaled divergence difference, $\Delta d_{Pn}(g, M_R M_A, \underline{x}) = 2n\{K(g, m_A^*, \underline{x}) - K(g, m_R^*, \underline{x})\} + c_n(p_A - p_R)$ (see Box 4 definition 22), and $Ev_L\left(M_R, M_A; \hat{g}_{n,\underline{x}}, \underline{x}\right) = E_{\hat{g}_{n,\underline{x}}}\left(2n\{K(\hat{g}_{n,\underline{x}}, m_{A,\underline{X}}^*, \underline{x}) - K(\hat{g}_{n,\underline{x}}, m_{R,\underline{X}}^*, \underline{x})\} - c_n(p_A - p_R)\right)$ (see Box 4 definition 24). The difference between the global and local is that in the calculation of the global evidence the observed data, $\underline{x}$, are considered as a realization of a random vector, $\underline{X}$, both in the estimation of the models to be compared and in the data on which they are compared. While in the local evidence, $\underline{x}$ is considered random in the estimation of the models but fixed as the data on which they are compared.

It is well established in statistics that providing an estimate of an unknown quantity is not sufficient; one must provide uncertainty associated with such an estimate. We use aleatory probability to quantify this uncertainty (Lele, 2020). In quantifying the pre-experiment uncertainty in evidence, we ask the question: How variable would the evidence be if we were to repeat the experiment? This is represented by the global (pre-experiment) sampling distribution of the evidence function. This distribution does not depend on the particular data set in hand.

When the competing models are fully specified and the reference model is the true model, Royall (1997, 2002) used the asymptotic Normal distribution of the LLR to approximate the sampling distribution of the evidence function and calculate the error probabilities. In Dennis et al. (2019), we derived the asymptotic distributions of the evidence function when the competing models are not fully specified and the true model is not part of the competing model spaces to approximate the sampling distribution and compute the error probabilities.

## 4.2 Uncertainty in evidence:

An important element common to all of our bootstrap procedures is that the complete evidence functions are the objects bootstrapped, not the component divergences. Thus, if the difference of information criterion values is the evidence function used, an such a bootstrap will produce a single distribution of $\Delta ICs$ rather than two distribution of $IC$ values. This is necessary because the geometry of model misspecification (Dennis et al., 2019; Ponciano and Taper, 2019, see also table 3) can create covariances (positive and negative) between the component divergences. These need to be captured by a bootstrap for it to accurately reflect the uncertainty in evidence. The non-parametric bootstrap method for the two cases described above is as follows.

### 4.2.1 Global uncertainty in evidence for the fully specified models:

Notice that in the bootstrap procedure in section 4.1.1, we are bootstrapping the difference in the log-likelihood jointly and not each component separately. Evidence, innately, is a comparison between two quantities. Clearly uncertainty in evidence involves not just the variances of each component but also covariance between them. The uncertainty reflected in the bootstrap distribution accounts for the covariance also. Thus, if the two models are positively correlated with each other, the uncertainty is reduced whereas if they are negatively correlated, the uncertainty is higher than the sum of variances. This, thus, takes into account the geometry of the model spaces appropriately, even when the models are fully specified. The quantiles of the smoothed bootstrap density of $Ev^{raw}\left(m_R, m_A; \hat{g}_{n,\underline{x}}, \underline{x}_b\right)$ give us confidence intervals for evidence (see Box 6 for an explicit algorithm).

### 4.2.2 Global uncertainty in evidence for model spaces with unknown parameters

Bootstrapping can also be used to obtain global confidence intervals for evidence with estimated parameters. The only difference is that the quantity bootstrapped is $Ev_G^{raw}(M_R, M_A; \hat{g}_{n,\underline{x}}, \underline{x}_b)$, which is in this paper a difference of information criterion values (see Box 5 for an explicit algorithm).

### 4.2.3 Local uncertainty in evidence

Lele (2020) reviewed the philosophical problems associated with global (pre-experiment) uncertainty and discussed the use of local (post-experiment) uncertainty in the context of linear regression. To recap, suppose we have only one covariate and we are fitting a linear regression through origin model. That is, the data are $(x_i, y_i), i = 1, 2, ..., n$ .and we fit the model $Y_i = \beta X_i + \grave{o}_i$ where $\grave{o}_i \sim N(0, \sigma^2)$ are independent, identically distributed random variables. The maximum likelihood estimator of $\beta$ is, $\hat{\beta} = \sum Y_i X_i / \sum X_i^2$ .

The question is: what is the variance of $\hat{\beta}$ ? If we consider the covariates to be random (this is the case when the experiment is not a designed experiment but an observational study), then $var(\hat{\beta}) = \sigma^2 E\left(1/\sum X_i^2\right)$. If $X_i \sim N(0,1)$, then $var(\hat{\beta}) = \sigma^2 / (n-2)$. This variance, which we term the global variance, is sometimes called an unconditional or pre-data variance. On the other hand, if we consider the covariates to be fixed, as is the case in designed experiments,

$var(\hat{\beta} \mid x_{1,} x_{2}, ..., x_{n}) = \sigma^{2} / \left\{ \sum x_{i}^{2} \right\}$. This variance, which we call the local variance, is sometimes called the conditional or post-data variance. Local uncertainties are appropriate only if the specified model is adequate. If the model is not adequate, these could be misleading.

The conditional variance is the variance most ecologists use when conducting regression analysis. Notice that conditional variance depends on the configuration of covariates the researcher observes in their particular data set. If the covariate values are highly dispersed, the slope is extremely well estimated; on the other hand, if the observed covariates values are not very different from each other, the slope is estimated with large uncertainty.

The local (conditional) variance makes intuitive sense: good data, strong inference; bad data, weak inference. It is argued (e.g. Goutis and Casella 1995) that the global (unconditional) inference does not reflect this differentiated inferential value of the observed data appropriately. Even if the researcher happens to have good, dispersed covariates, the global variance does not recognize that and increases the variance because the researcher, in another replication of the experiment could have observed less dispersed covariates and vice versa. We note that the pairwise resampling used in bootstrap inference for regression gives the unconditional variance even when the model is misspecified.

For local uncertainty, the sample space over which the variation is considered is a subset of the total sample space. This is called a 'relevant subset' (Buehler, 1959). Such a relevant subset is often determined using an ancillary statistic. An ancillary statistic is a function of the data whose distribution does not depend on the parameters. There are, often, multiple ancillary statistics (Basu, 1964; Pena et al., 1992) and hence relevant subsets are not necessarily unique. In our opinion, the appropriateness of the relevant subset is determined based on the type of future experimental replication one envisions. Different future experiments determine different relevant subsets as was the case in the Mark-Capture-Recapture example in Box 2.

It has been argued that local (post-experiment, post-data, conditional) confidence intervals are preferable as the measure of uncertainty because they reflect the informativeness of the data at hand appropriately. If the data are highly informative, the local confidence intervals are shorter than the global confidence intervals and if the data are not informative, the local confidence intervals appropriately are wider than the global confidence intervals. Again, this argument hinges on the model being correctly specified.

Some august statisticians (e.g. Royall, 2004) argue the local interval is the only one that should be used irrespective of the design because design is an ancillary statistic and has no impact on the inference once the data are obtained. If the data are highly informative either by design or by chance, we should be quite confident about our estimate of the total population size, irrespective of what other experimenters might observe. It can be shown (see review in Lele, 2020) that prediction based on local uncertainty is more accurate than predication based on global uncertainty. However, this result also depends on correct model specification.

On the other hand, other equally august statisticians (e.g. Cox 2004 in his discussion of Royall (2004)) claim design should play a role in uncertainty quantification.  We agree with this latter opinion on the importance of design. Both because the interpretation of uncertainty intervals should depend on the potential type of the future experimental replication, and thus so should the choice of the ancillary statistics or relevant subsets.  And because, as we show in section 5.2, the accuracy of the local interval depends on correct model specification to a greater degree than does the global.

### 4.2.4   Local uncertainty when comparing two model spaces:

Local evidence uncertainty in the comparison of model spaces is calculated similarly to global evidence uncertainty. Data sets are repeatedly reconstructed by bootstrapping the original data. With each bootstrapped data set model parameters for both reference and alternative models are reestimated and an evidence value comparing the models is calculated. The critical distinction between global and local uncertainty is that in the local calculations the likelihood for each bootstrapped model is evaluated using the original data not the bootstrapped data (see Box 6 for an explicit algorithm).

In section 5, we use simulations to study the coverage properties of the global and local sampling distributions. Both the cases of linear regression and structural equation models are investigated.

---

Box 6: Bootstrap algorithms for global and local evidence uncertainty.

All of the bootstraps described in this box can be performed using the R function KKICv, which we supply in the supplemental material.

Evidence uncertainty for specified models:

1) Obtain a random sample of size $n$ with replacement from the original sample. This bootstrap sample is denoted by $\underline{x}_b = (x_{b1}, x_{b2}, ..., x_{bn})$.

2) Evaluate the evidence at the bootstrap sample, namely, $Ev^{raw}\left(m_R, m_A; \hat{g}_{n,\underline{x}}, \underline{x}_b\right) = -2\left(l_{m_A}\left(\underline{x}_b\right) - l_{m_R}\left(\underline{x}_b\right)\right)$.

3) Repeat steps 1 and 2 B times and accumulate to get the set of results $\{Ev^{raw}(m_R, m_A; \hat{g}_{n,\underline{x}}, \underline{x}_b), b = 1, 2, ..., B\}$.

4) Estimate the density function of the $\{Ev^{raw}(m_R, m_A; \hat{g}_{n,\underline{x}}, \underline{x}_b)\}$ in 3).

5) Calculate $Ev(m_R, m_A; \hat{g}_{n,\underline{x}}, \underline{x})$ as the expectation (mean) of the estimated density from step 4.

Global evidence uncertainty estimation:

1) Obtain a simple random sample of size n with replacement from the observed data $\underline{x}$. Let us denote this by $\underline{x}_b = (x_{b,1}, x_{b,2}, ..., x_{b,n})$.

2) Based on this bootstrap data, estimate the model parameters for each model space. Let us denote these models by $\hat{m}_{R,b}$ and $\hat{m}_{A,b}$. These are projections of the empirical CDF of the bootstrap data onto the corresponding model spaces.

3) Compute and store $Ev_G^{raw}(M_R, M_A; \hat{g}_{n,\underline{x}}, \underline{x}_b) = -2\{l_{\hat{m}_A, \underline{x}_b}(\underline{x}_b) - l_{\hat{m}_R, \underline{x}_b}(\underline{x}_b)\} + c_n(p_A - p_R)$. The smoothed density of $Ev_G^{raw}(M_R, M_A; \hat{g}_{n,\underline{x}}, \underline{x}_b), b = 1, 2, ..., B$ is the bootstrap estimate of the sampling distribution of $Ev_G^{raw}(M_R, M_A; \hat{g}_{n,\underline{x}}, \underline{x}_b)$.

4) Quantiles of the smoothed density of $Ev_G^{raw}(M_R, M_A; \hat{g}_{n,\underline{x}}, \underline{x}_b)$ give us confidence intervals for evidence.

5) Calculate
$Ev_G(M_R, M_A; \hat{g}_{n,\underline{x}}, \underline{x})$ as the expectation (mean) of the estimated density from step 4.

**Local evidence uncertainty estimation:**

1) Generate a random sample with replacement and of size $n$ from the observed data. Let us denote this by $\underline{x}_b = (x_{b,1}, x_{b,2}, ..., x_{b,n})$.

2) Re-estimate the parameters using the bootstrap sample. Let us denote them by $\hat{\theta}_{R,b}$ and $\hat{\theta}_{A,b}$.

3) Compute $Ev_L^{raw}(M_R, M_A; \hat{g}_{\underline{x}_b}\underline{x}) = -2\{l_{\hat{m}_A, \underline{x}_b}(\underline{x}) - l_{\hat{m}_R, \underline{x}_b}(\underline{x})\} + c_n(p_A - p_R)$.

4) Use the quantiles of the smoothed bootstrap distribution of $Ev_L^{raw}(M_R, M_A; \hat{g}_{\underline{x}_b}\underline{x})$ to quantify conditional uncertainty of the strength of local evidence.

5) Calculate $Ev_L(M_R, M_A; \hat{g}_{n,\underline{x}}, \underline{x})$ as the expectation (mean) of the estimated density from step 4

__________________________End of Box 6________________________________________

# 5    Simulation Validation.

If new statistical approaches are proposed, the scientific community has a legitimate expectation that they will be validated both mathematically, and computationally (Devezer et al., 2020). For a procedure that generates confidence intervals, whether global or local, to be a legitimate frequentist procedure, they need to cover/capture their targets at least the specified level (Casella, 1992). The fundamental difference between global and local inference is that a global target cannot depend on the data at hand, while a local target must depend on the data at hand.

Globally we want our intervals to cover the global penalized scaled divergence difference:
$\Delta D_{Pn}(g, M_R, M_A) = 2n\{K(g, M_A) - K(g, M_R)\} + c_n(p_A - p_R)$. Locally we want our intervals to cover the local penalized scaled divergence difference:
$\Delta d_{Pn}(g, M_R, M_A, \underline{x}) = 2n\{K(g, m_A^*, \underline{x}) - K(g, m_R^*, \underline{x})\} + c_n(p_A - p_R)$. For the Kullback-Leibler divergence, this is simply $-2\{l(m_A^*; \underline{x}) - \log m_R^*(x_i)\} + \log(n)(p_R - p_A)$, the penalized scaled LLR for the observed data under the best approximating models in the two competing spaces to the true generating mechanism. We note this is identical to what is considered the target likelihood in the general profile likelihood literature, e.g. section 3.1 of Pace and Salvan (2006).

## 5.1    Global and local coverages in alternate model space topologies

There are 14 possible topologies for a reference model space, an alternative model space and a generating process.  The model spaces compared can be nested, overlapping, or disjoint.  If the model comparison is correctly specified, the generating process will be in at least one of the model spaces. If the comparison is misspecified then the generating process will be in neither model space. Table 3 describes coverage and interval length for the global and local confidence intervals of the strength of evidence for model comparisons in each of these topologies in a simple multiple regression example (see the table legend for simulation details).

Table 3, The behavior of our global and local uncertainty procedures in all 14 possible model specification topologies. In the g location figures the solid ellipse indicates the reference model space while dashed ellipse indicate the alternative model space.  For the correctly specified comparisons, cases 1-7, the star indicates the location of the generating process.  For the misspecified comparisons, the arrow indicates the location of the projection from the generating process to the model spaces. The asymptotic distribution refers to the unpenalized likelihood ratio statistic (often denoted G2); the penalty term for converting G2 to an evidence function produces location-shifted versions of the asymptotic distributions (Dennis et al., 2019). The covariates are three N(0,1) random vectors and are held constant over all simulations. For each line, the coefficients in the generating model of the three covariates (there are no interactions) are given in the column g par. In all simulations the intercept is 2.0 and the error standard deviation is 1. The sample size for all simulations in this table is 100, a realistic size for ecological studies, and one that meets most common rules of thumb for multiple regression.  Coverage proportions were estimated using 1000 trials for each case.  Coverage is reported for nominal 95% and 90% kde1d intervals.  Mean interval length and its standard deviation is also reported.

| Case | g location | Asymptotic Distribution | Exemplar | g par | global coverage 95%/90% | Length mean (sd) 95%/90% | local coverage 95%/90% | Length mean, sd 95%/90% |
|---|---|---|---|---|---|---|---|---|
| 1 | 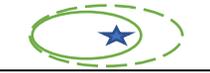 | chisquare | $Ev(g=m_{001}, M_R=M_{001}, M_A=M_{011}; x)$ | 0.00 0.00 0.15 | 0.00 0.00 | 8.18 (3.67) 6.48 (3.16) | 0.99 0.97 | 6.66 (1.17) 4.90 (0.86) |
| 2 | 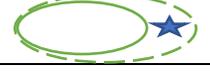 | Non-central chisquare | $Ev(g=m_{011}, M_R=M_{001}, M_A=M_{011}; x)$ | 0.00 0.30 0.15 | 0.95 0.88 | 22.79 (7.42) 19.06 (6.29) | 0.98 0.95 | 8.12 (1.39) 6.03 (1.00) |
| 3 | 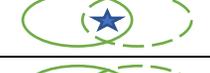 | Weighted sum of chisquare | $Ev(g=m_{010}, M_R=M_{110}, M_A=M_{011}; x)$ | 0.00 0.30 0.00 | 1.00 1.00 | 13.75 (4.03) 10.58 (3.41) | 0.99 0.97 | 10.94 (1.37) 7.84 (0.95) |
| 4 | 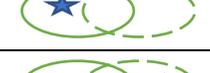 | Normal | $Ev(g=m_{110}, M_R=M_{110}, M_A=M_{011}; x)$ | 0.60 0.30 0.00 | 0.95 0.90 | 44.89 (5.91) 37.62 (4.97) | 0.98 0.94 | 13.29 (1.4) 9.90 (0.97) |
| 5 | 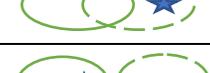 | Normal | $Ev(g=m_{011}, M_R=M_{110}, M_A=M_{011}; x)$ | 0.00 0.30 0.15 | 0.98 0.93 | 18.23 (5.71) 14.51 (4.96) | 0.98 0.96 | 11.12 (1.35) 8.02 (0.95) |
| 6 | 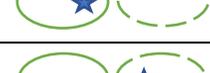 | Normal | $Ev(g=m_{110}, M_R=M_{110}, M_A=M_{001}; x)$ | 0.60 0.30 0.00 | 0.96 0.92 | 48.00 (5.53) 40.25 (4.66) | 0.97 0.93 | 14.23 (1.42) 10.69 (1.01) |
| 7 | 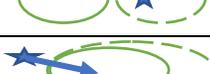 | Normal | $Ev(g=m_{001}, M_R=M_{110}, M_A=M_{001}; x)$ | 0.00 0.00 0.15 | 0.95 0.85 | 20.67 (5.55) 16.56 (4.78) | 0.99 0.96 | 12.9 (1.36) 9.53 (0.98) |
| 8 | 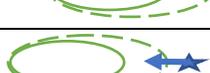 | Weighted sum of chisquare | $Ev(g=m_{111}, M_R=M_{001}, M_A=M_{011}; x)$ | 0.05 0.05 0.15 | 0.00 0.00 | 8.11 (3.58) 6.42 (3.09) | 0.97 0.93 | 6.73 (1.18) 4.95 (0.85) |
| 9 | 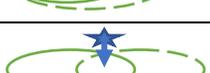 | Normal | $Ev(g=m_{111}, M_R=M_{001}, M_A=M_{011}; x)$ | 0.05 0.30 0.15 | 0.94 0.88 | 22.73 (7.12) 19.01 (6.02) | 0.99 0.97 | 8.08 (1.36) 6.01 (0.99) |
| 10 | 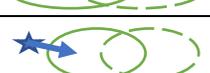 | Weighted sum of chisquare | $Ev(g=m_{111}, M_R=M_{110}, M_A=M_{011}; x)$ | 0.05 0.30 0.05 | 0.99 0.98 | 15.1 (4.69) 11.75 (4.02) | 0.97 0.93 | 10.92 (1.41) 7.84 (0.94) |
| 11 | 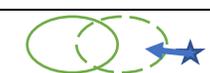 | Normal | $Ev(g=m_{111}, M_R=M_{110}, M_A=M_{011}; x)$ | 0.60 0.30 0.05 | 0.96 0.90 | 45.47 (6.09) 38.09 (5.12) | 0.99 0.97 | 13.38 (1.42) 9.98 (1.01) |
| 12 | 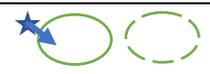 | Normal | $Ev(g=m_{111}, M_R=M_{110}, M_A=M_{011}; x)$ | 0.05 0.30 0.15 | 0.99 0.96 | 18.98 (5.88) 15.14 (5.09) | 0.98 0.94 | 11.08 (1.37) 8.01 (0.98) |
| 13 | 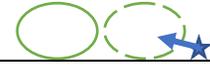 | Normal | $Ev(g=m_{111}, M_R=M_{110}, M_A=M_{001}; x)$ | 0.60 0.30 0.05 | 0.95 0.92 | 49.05 (5.9) 41.1 (4.97) | 0.98 0.96 | 14.33 (1.44) 10.77 (1.01) |
| 14 | 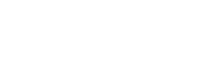 | Normal | $Ev(g=m_{111}, M_R=M_{110}, M_A=M_{001}; x)$ | 0.05 0.05 0.15 | 0.95 0.88 | 22.55 (5.6) 18.18 (4.8) | 0.97 0.94 | 12.93 (1.36) 9.56 (0.95) |

A number of interesting patterns can be observed in Table 3. In 12 of the 14 possible model space topologies, the global intervals cover reasonably, with actual coverages close to nominal coverages. Cases 1 and 8, however, have no coverage! Case 1 is the topology of nested models with the generating process in the reduced model. The asymptotic distribution for this case is chisquare. Case 8 represents the misspecified analog of Case 1, the approximating models are nested with the generating process closest to the reduced model. The asymptotic distribution for case 8 is a weighted sum of chisquare. This is a very flexible distribution, and in this case generates a distribution indistinguishable from a chisquare distribution. Alarm at this complete lack of coverage in these two cases is somewhat reduced by recognizing that the target ($\Delta D_{P_n}(g, M_{R}, M_{A})$) is the boundary of these chisquare distributions.

On the other hand, the local confidence intervals for evidence behave well in all 14 possible model space topologies. In all cases local interval coverage exceeds the nominal levels. Overcovering is acceptable in approximate confidence intervals, particularly if interval length is narrow. In all cases of Table 1, the average lengths of the local intervals are less than that of global intervals. This is not always the case. For very small sample size, the average local interval length may exceed the average global interval length (see Figure 6).

## 5.2    Sample size and interval lengths.

In the linear models example of Table 3, global and local intervals respond quite differently to changes in sample size. These differences are explored in Figures 6 and 8. Figure 6, panel A shows box plots of the ratio of local interval length to global interval length over a range of increasing sample size for the case of case 1 from Table 1. The models compared are nested and the generating process is in the reduced model. The asymptotic distribution of evidence is chisquared. At lower sample sizes the local interval length generally exceeds the global length. At higher sample sizes the local interval is generally shorter than the global interval, with the ratio appearing to approach a limit of at about 0.6. Model topologies shown in case 1 and 8 of Table 3 behave in this fashion.

Figure 6 panel B represents case 4 from Table 1. The models compared are overlapping with the generating process located in the non-overlapping portion of the reference model. The interval length behavior here is very different from that in panel A. Local intervals exceed global intervals only at the smallest sample sizes. Further, the local/global interval length ratio rapidly decreases towards 0 (rate $1/\sqrt{n}$). All model topologies except those of cases 1 and 8 behave in this fashion.

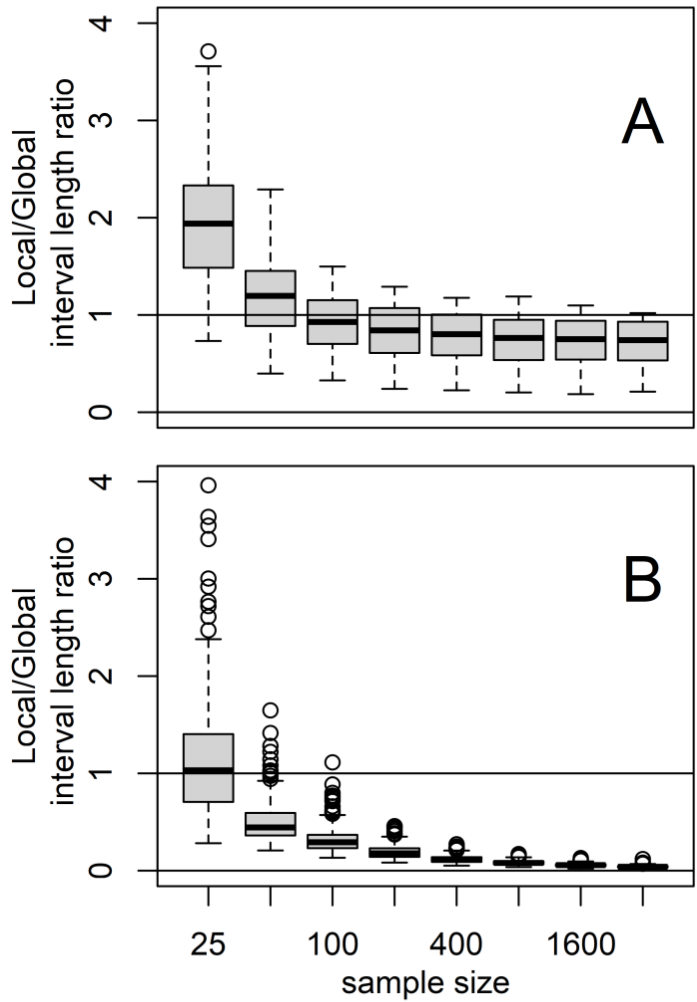

Figure 6: Box plots of the ratios of the local and global interval lengths as a function sample size. Each box summarizes the results for 1000 simulations. In panel A, all parameters (except for sample size) are set to those case 1 in Table 3. In panel B, all parameters (except for sample size) are set to those of case 4 in Table 3.

These differences in global to local interval length in relation to sample size coupled with the fact that mean evidence grows linearly with sample size in both cases have considerable impact on inference and experimental design. The ability of the global interval to distinguish the observed evidence from 0 grows very slowly with sample size. On the other hand, the local interval will be able to detect real difference from 0 or either of our two thresholds with relatively small sample sizes.

**5.3   Model set misspecification and evidential uncertainty.**

Here we demonstrate the effect of model set misspecification on the uncertainty of evidence with simulations based on the Grace and Keeley example. We look at 4 different conditions of model set adequacy: A) correctly specified comparison with very strong evidence, B) correctly specified comparison with strong evidence, C) a mildly misspecified model comparison, and D) a badly misspecified model comparison.

In case A), we compare the model that is the GKBM without the weakest path (GKBM − R~L ) with a model that is the GKBM without the second weakest path (GKBM − R~C ). The data in these simulations are generated from the estimated (GKBM − R~L). The generating process is in the compared model set; therefore, the comparison is correctly specified.

In case B) we estimate and compare the same models as in case A. The generating model has same form as in case A (all the same paths are present) but one of the coefficients (R~P) has been weakened from 0.299 to 0.205. The model set is still correctly specified, but the penalize divergence differences (whether global or local, see definitions 19 and 20) between the compared models is less than in case A. Consequently, the distribution of realized evidences (definitions 23 and 24) will be shifted to lower values.

Case C) compares the same models as in case A) {GKBM − R~L, GKBM − R~C}. The data are generated by the GKBM. Since the generating process (GKBM) is quite close to one of the models in the model set (GKBM – R~L), the comparison is only mildly misspecified.

Finally, in case D) we compare a model that is the GKBM without the second strongest path (GKBM − C ~ L) with a model that is the GKBM without the strongest path (GKBM − F ~ A). As in case B), the data are generated by the GKBM. Since the generating process (GKBM) is quite different from both of the models in the model set, the comparison is badly misspecified.

Table 4 indicates that, at least in this example, under correct model specification, a researcher is very unlikely to obtain secure misleading evidence using either interval. On the other hand, the researcher is more likely to correctly obtain strong and secure evidence using the conditional interval than with the unconditional interval. If the model set is misspecified, secure misleading evidence becomes a possibility, and much more so using the conditional interval than the unconditional interval.

Interestingly, the average reliability (proportion of the time correct model is identified) is always slightly greater using the local evidence distribution rather than when using the global evidence distribution. This agrees with the previous results (Aitchison1975; Royall and Cumberland, 1985; and Vidoni,1995) that indicate predictive accuracy is greater using conditional inference.

| Case | Model Set Adequacy | Interval Type | Evidential Security Categories | | | | | | | | | Average Reliability |
|------|--------------------|---------------|------|------|-------|-------|-------|-------|-------|------|-------|---------------------|
|      |                    |               | MS   | CS   | MI    | CI    | W     | PI    | SI    | PS   | SS    |                     |
| A    | Correctly Specified | Global       | 0    | 0    | 0.001 | 0     | 0.069 | 0.108 | 0.447 | 0    | 0.375 | 0.944               |
|      |                    | Local         | 0    | 0    | 0     | 0.001 | 0.069 | 0.105 | 0.101 | 0    | 0.724 | 0.975               |
| B    | Correctly Specified | Global       | 0    | 0    | 0.001 | 0.003 | 0.345 | 0.195 | 0.371 | 0    | 0.085 | 0.834               |
|      |                    | Local         | 0.001| 0    | 0     | 0.003 | 0.336 | 0.202 | 0.152 | 0    | 0.306 | 0.877               |
| C    | Mildly Mis-Specified | Global      | 0.003| 0    | 0.042 | 0.066 | 0.260 | 0.140 | 0.390 | 0    | 0.099 | 0.720               |
|      |                    | Local         | 0.034| 0    | 0.012 | 0.063 | 0.256 | 0.148 | 0.126 | 0    | 0.361 | 0.775               |
| D    | Badly Mis-Specified | Global       | 0.003| 0    | 0.068 | 0.050 | 0.261 | 0.114 | 0.400 | 0    | 0.104 | 0.711               |
|      |                    | Local         | 0.046| 0    | 0.025 | 0.050 | 0.260 | 0.115 | 0.137 | 0    | 0.367 | 0.761               |

Table 4: Models compared and generating process for each model set are described in the text. The bootstrap mean evidence is used as the strength of evidence. Each row lists the proportions each security category occurs in 1000 simulations and the overall reliability. Security in each row is determined either by the unconditional evidential confidence intervals or the conditional evidential confidence intervals. The categories of security are: MS (misleading and secure), CS (confusing and secure), MI (misleading and insecure), W (weak), PI (prognostic and insecure), SI, (strong and insecure), PS (prognostic and secure), and SS (strong and secure). Reliability is the proportion of times the best model is correctly identified—by any strength of evidence—averaged over all trials.

The table gives the impression that there is little difference between mildly and badly misspecified model sets in regard to evidence. But this is only because the choice of the mean of the smoothed bootstrapped $\Delta$SIC as the measure of the strength of evidence rather than the raw $\Delta$SIC has profound impact. Figure 7 presents the same data used to calculate table 4 in another fashion. Here both the global and local intervals are explicitly plotted for each 1000 trials in the simulations of cases A, B, C and D. The trials are sorted along the x axis by the mean smoothed bootstrapped strength of evidence. Panel E plots the same simulations and intervals as panel D, however, in this case the trials are sorted by the raw $\Delta$SIC —not by the mean smoothed bootstrapped $\Delta$SIC . We do not show plots with similar reordering for panels A, B, and C because in these cases the differences between the raw $\Delta$SIC and the mean of the smoothed bootstrap are not visually perceptible.

In cases A, B, and C the difference between raw $\Delta$SIC and smoothed mean bootstrapped $\Delta$SIC are quite small and the correlation of raw $\Delta$SIC and mean smoothed bootstrapped $\Delta$SIC are greater than 0.99. Thus, there is almost no impact of choice of evidence measure in these cases with correct and mild misspecification. In the badly misspecified case D, there is a large average difference between raw $\Delta$SIC and the mean smoothed bootstrapped $\Delta$SIC and almost no correlation between them. Further, when using the raw $\Delta$SIC , the location of the security intervals becomes almost unrelated to the strength of evidence. Consequently, the raw $\Delta$SIC has almost no ability to securely identify the best model.

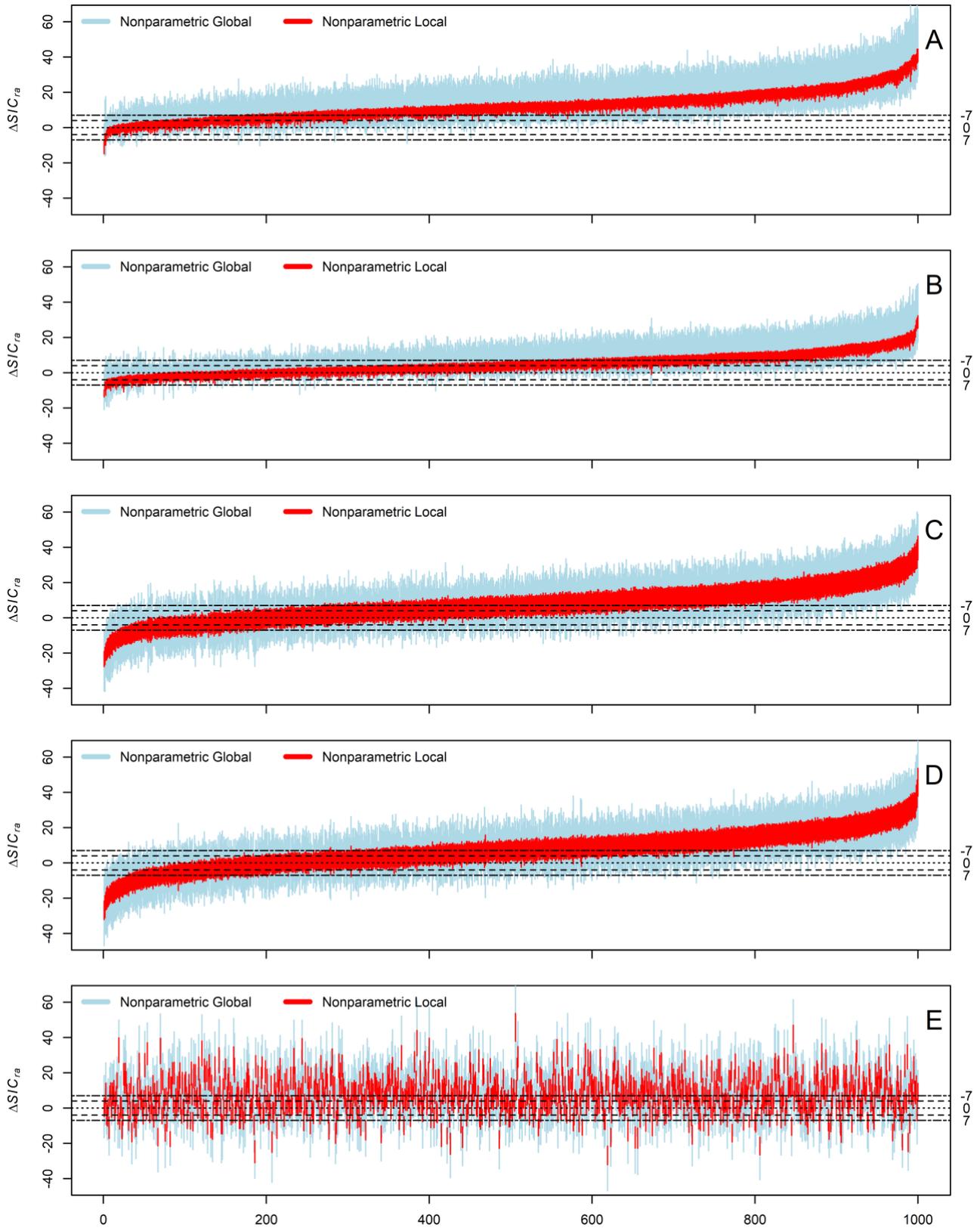

Figure 7: Global (blue) and local (red) 90% confidence intervals for the 1000 simulations for cases A, B, C, and D, described in the text. In panels A-D simulations are ordered by mean smoothed local

evidence. Panel E presents the same data as panel D but ordered by the raw evidence. Panels A-C not shown reordered because with a Spearman correlation of >=0.997 between raw and mean smoothed evidence in these cases, there is no perceptible change in the figures.

## 6    Discussion

Historically, the appeal of classical Neyman-Pearson testing has been the appearance of a strong control of error probabilities. Dennis et al. (2019) show this to be an illusion. Under model misspecification, the realized error rate for a NP test can be less than or greater than its nominal rate. In some realistic cases the probability of error in a NP test can even increase with increasing sample size. Evidential analysis is superior to NP testing in that the total error rate always decreases with increasing sample size, both under correct model specification and under model misspecification.

However, as Dennis et al. (2019) further points out that evidence is not entirely immune to problems due to model misspecification. Under misspecification, the probability of strong misleading evidence is not directly calculable because the generating process is not one of the models compared and is not even known. This current paper demonstrates that evidential error rates can be estimated even under model misspecification using non-parametric bootstrapping techniques (at least for independent data). Our approach to the bootstrapping of evidence differs from that used in the EIC (Konishi and Kitagawa, 1996; Ishiguro et al., 1997) in that we bootstrap the evidential comparison as a unit (see definitions 23 and 24 Box4) whereas the EIC compares bootstrapped components. The joint bootstrapping allows us to estimate the impact of model set misspecification on evidential uncertainty more effectively. In this paper, we have only addressed the case of independently distributed data. We expect, however, that this approach can be extended to other data structures with the use of subtler bootstrapping methods (Lele, 1991, 2003; Lahiri, 2003).

It is important for scientists seeking to use and interpret these measures of uncertainty to understand the two intervals, global and local, are quantifying two different kinds of uncertainty. Statistical evidence is an estimate of the relationship between two models and the generating process. It is a penalized sample size scaled estimate of the difference of the divergences of two models from the generating process (truth). Valid confidence intervals of an estimate tell us how confident we are that the estimation target lies within the interval. In the global case, our estimate of evidence is the mean of the global bootstrap distribution of evidence, but the estimation target is the true penalized scaled divergence difference (Box 4, definition 19). In the local case our estimate of evidence is the mean of the local bootstrap distribution of evidence, but the target is the true evidence in the data without model estimation error (Box 4, definition 20).

Our simulations show that both interval types have good coverage of their targets for correct and mildly misspecified model comparisons. For badly misspecified model comparisons local inference can be overconfident. Nevertheless, we are in a position to make scientific inferences about the true relationships of our compared models to the generating process, backed by an uncertainty measure warrant.

Both the global and local evidence confidence intervals are important to science because they answer different questions. The local uncertainty tells you how confident you are in your evidence given the data you have collected. On the other hand, the global interval is a confidence interval on the true penalized scaled divergence difference. This speaks directly to the relative ability of our models to represent nature. The resampling is non-parametric to accommodate model misspecification.

Further, the intervals incorporate both sample and model estimation uncertainty. Replication is the backbone of science as a social activity. The global uncertainty we offer answers the question of how dissimilar to the current evidence we would expect new evidence to be if our experiment were to be repeated. This is the interval that other researchers should consider when trying to decide if their new results call the current results into question.

Which interval should a scientist use? Unfortunately, a univocal recommendation is not possible. The local interval is tremendously appealing because it is so short and because its overall reliability is greater, but to justify inference based on it alone, the scientist needs to be able to defend the assumption of correct model set specification. In the rough and tumble world of ecology this will rarely be possible, except for tightly controlled experiments with well understood error structures. The global interval presents a more honest appraisal of the replicability of the scientist's results. If the global interval has been presented, the local interval can be a useful indication of how good the results could possibly be. For the accumulation of understanding through science, our personal preference is the global interval. This preference is grounded in our opening quote from Plato. Using the global interval you will accept wrong statements less frequently than when using the local interval. However, in a decision context, where costs and benefits are explicit, the local inference's property of making correct predictions more often than global inference might be important.

Hopefully, our recommendation to focus on the global interval will be only temporary. We expect that often model sets could be misspecified, but close enough to correctly specified that the local interval would be a justifiable improvement over the global interval. Research into diagnostics to identify these cases is called for. Useful diagnostics will involve more than measures of the adequacy of single models (e.g. Markatou and Sofikitou, 2019) they must somehow include measures of the geometry of the generating process and the competing models (Dennis et al., 2019; Ponciano and Taper, 2019)

In the meantime, little is practically lost. We agree with Goutis and Casella (1995) that "In any experiment both pre-data inferences and post-data inferences are important." Our inferential strategy is a hybrid of local and global (conditional and unconditional). Our primary tool is the strength of evidence this is local (i.e. conditional). The evidence expresses clearly what the data we says about the relationships among nature and our models. Our secondary tool is our pair of measures of the security of the evidence. If we choose a global (that is unconditional measure) we gain an honest, if perhaps overly conservative, insight into the degree that chance, experimental/sample design, and model misspecification may have influenced our evidence. If we choose a local (that is a conditional measure) we gain a more precise understanding of the information in the data, at the risk of overconfidence due to model misspecification. Much of statistics both classical and Bayesian relies on conditional inference and thus might be over-confident in its conclusions in the face of potential model misspecification (see also Yang and Zhu, 2018).

While the global uncertainty, either calculated from asymptotic theory or from the non-parametric bootstrap is a useful statistic, it should not be interpreted too literally. As Fisher (1945a, 1945b, 1955, 1956, 1960) long argued (see Rubin, 2020 for a detailed discussion) an exact repetition of an experiment is not possible in many branches of science. Certainly, this is true in ecology and environmental science, where heterogeneity and temporal data abound. To paraphrase Heraclitus, you can't electrofish the same river twice. A more realistic understanding of global uncertainty would come a metanalysis of the actual repetition of modestly sized experiments distributed in space and time than from a single large experiment. As an example, Jerde et al. (2019) conducts an evidential comparison of models for the intra-specific allometry of metabolic rate in fish using a database of 25 high quality studies, with 55 independent trials, across 16 fish species.

Jerde et al. (2019) use evidential support intervals in their analysis of the allometry of metabolic rate in fish. We wish to point out that, while both are useful, evidential support intervals and confidence intervals for evidence are different. Evidential support intervals indicate the range of parameter values in a model space that are not differentiated from the best estimate at a specified strength of evidence. Confidence intervals make a statement that at the specified probability a random interval, whose randomness stems from sample space probabilities, contains the true parameter value (Dennis, 2004). Under correct model specification, the support interval indicates over what range of parameter values the relative plausibility of the best estimate relative to the parameter value is less than the designated strength of evidence.

Under a correct model assumption, a $\Delta$AIC interval is directly transformable into a confidence interval using Wilks-Wald hypothesis test inversion (see Dennis et al., 2019). The confidence level of this transformed interval will depend only on the chosen strong evidence threshold, $k_R$. On the other hand, the level of a confidence interval corresponding to a $\Delta SIC$ interval will be a function of both $k_R$ and $\log(n)$. As $n$ increases the confidence level will increase. A confidence interval is preferred if a true model assumption is justified. Using a support interval rather than a confidence interval acknowledges that your model set may be misspecified. Here, a support interval is a more honest, if weaker statement.

Confidence intervals for the strength of evidence, at least as we have developed them in this paper, are used in the comparisons of model spaces. The interval is not on the space of parameter values but on the space of strength of evidence values. For problems with data structures that will allow bootstrapping, the global confidence intervals for evidence gives an estimate of the frequentist confidence that the "true penalized scaled divergence difference" is in the interval—even in the presence of model misspecification.

## 7    Conclusion

Neither the Bayesian nor classical frequentist statistical toolkits appear adequate for the increasingly complex challenges of the future. In the long run, neither our models nor our data, nor our conclusions are static. We need to look at multiple models realizing that we do not know truth and evolve these models towards better approximations of truth with the accumulation of data and use of evidence as a selection function.

We have produced both global and local uncertainty measures that are easily calculated for many analyses using the R-code that we supply in the supplement. Further, by creating three categories for the strength of evidence coupled with three categories for the security of evidence we have constructed a conceptual language that allows scientists a statistically valid way to talk, and publish, about interesting results that are not yet conclusive.

## 9    Conflict of Interest

All authors declare that the research was conducted in the absence of any commercial or financial relationships that could be construed as a potential conflict of interest.

## 10    Author Contributions

All authors conceived of this study jointly.  R Code written my MLT.  MLT & SRL jointly wrote the first draft.  All authors contributed to the many draft revisions.

## 11    Funding

All authors were partially funded by a grant from the Japan Society for the Promotion of Science to professor Shimatani, Kenichiro. J.M. Ponciano was partially supported by two grants from the US National Institute of Health grant number to University of Florida NIH R01 GM103604 to J.M. Ponciano and NIH 1R01GM117617 to professors Jason K. Blackburn and J.M. Ponciano

## 12    Acknowledgments

This paper is in review for the journal Frontiers in Ecology and Evolution. The journal has kindly permitted archiving of the submission version.  Changes can be expected due to review.

## 13    Data Availability

The data for the Grace and Keeley (2006) structural equation model can be found as the data for SEM.6 at < https://www.usgs.gov/centers/wetland-and-aquatic-research-center/science/quantitative-analysis-using-structural-equation?qt-science_center_objects=0#qt-science_center_objects >. The data can also be found as GK.d.rds in the supplementary material

## 14    Supplementary Material

The files KKICv.R, KKICv.demo.R, GK.d.rds, and README.md can be found at <https://github.com/jmponciano/mltaper-bootstrap>.